\documentclass[aps,prb,floatfix,longbibliography,superscriptaddress,preprint,showkeys]{revtex4-1}

\usepackage[english]{babel}

\usepackage[utf8]{inputenc}
\usepackage[T1]{fontenc}

\usepackage{amsthm}
\usepackage{amssymb}
\usepackage{mathtools}
\usepackage{units}
\usepackage{braket}

\usepackage{xcolor}
\usepackage[pdftex]{graphicx}

\usepackage{hyperref}
\hypersetup{colorlinks=true, citecolor=blue, urlcolor=blue, linkcolor=blue}

\newcommand{\operator}[1]{\hat{#1}}

\newcommand{\euler}{\mathrm{e}}
\newcommand{\cone}{\mathrm{i}}
\renewcommand{\vec}[1]{\boldsymbol{#1}}
\newcommand{\code}[1]{\textsf{#1}}

\renewcommand{\Re}{\mathrm{Re}}
\renewcommand{\Im}{\mathrm{Im}}

\newcounter{exercise} 
\renewcommand{\theexercise}{\arabic{exercise}}
\newenvironment{exercise}[1][]{
  \refstepcounter{exercise}
  {
  \par\noindent\rule{\linewidth}{0.4pt}\par  
  }
  \par\noindent\textbf{\textsf{Exercise~\theexercise}}%
  \ifx\relax#1\relax\else~(\textsf{#1})\fi
  \par\sffamily
}
{
  \par\noindent\rule{\linewidth}{0.4pt}\par  
  }

\begin{document}
\title{Exploring Quantum Entanglement: A Hands-on Course on Spin Dynamics and Entanglement Transfer with Python Modules}

\author{Salomo Cedric Karst}
\affiliation{Landesschule Pforta, Schulpforte, 06628 Naumburg -- Bad Kösen, Germany}
\email[Correspondence e-mail address: ]{salomo.karst@student.uni-halle.de, salomokarst19@gmail.com}

\author{J{\"u}rgen Henk}
\affiliation{Institut f{\"u}r Physik, Martin-Luther-Universität Halle-Wittenberg, 06099 Halle, Germany}
\email[Correspondence e-mail address: ]{juergen.henk@physik.uni-halle.de}

\date{\today}

\begin{abstract}
Quantum entanglement is a captivating phenomenon in quantum physics, characterized by intricate and non-classical correlations between particles. This phenomenon plays a crucial role in quantum computing and measurement processes. In this hands-on course students explore the dynamics of quantum systems with up to three spins, introducing the fundamental mechanisms by which entanglement evolves and transfers within such systems. Through detailed examples, simulations, and analyses, the hands-on course offers insights into the fundamental principles of entanglement. The provided \code{python} modules reproduce  the presented results and serve as a basis for further projects. The target audience of this hands-on course is physics enthusiasts among students in their first semesters.
\end{abstract}


\maketitle

\section{Introduction}
\label{sec:introduction} 
Quantum entanglement is a fundamental phenomenon where particles become so deeply correlated that their quantum states cannot be described independently, even across large distances. First discussed by Einstein, Podolsky, and Rosen in 1935~\cite{Einstein1935}, entanglement challenges classical intuitions and has far-reaching implications for our understanding of reality~\cite{Lewis2016, Susskind2014}. Experiments by Aspect, Clauser, and Zeilinger, recognized by the 2022 Nobel Prize~\cite{Nobel2022}, confirmed its nonlocal nature and manipulability.

Entanglement features
\begin{itemize}
    \item Non-local correlations: measurements on one particle instantaneously affect its entangled partner.
    \item Measurement effects: the act of measurement collapses the superposition for both particles.
    \item Quantum superposition: entangled systems exist in a combined state until observed.
\end{itemize}
These properties underpin applications in quantum computing, cryptography, and teleportation~\cite{Bayat2022}.  Entanglement also plays a role in quantum decoherence, which explains the transition from quantum to classical behavior via environmental interactions~\cite{Schlosshauer2007}. Entanglement swapping is one mechanism by which quantum information transfers between non-interacting particles.

This hands-on course offers a pedagogical study of entanglement and its transfer, using qubits as model systems, following Susskind’s approach~\cite{Susskind2014, theoretical-minimum}. Qubits, unlike classical bits, can exist in superpositions and become entangled. Represented as spin-$\nicefrac{1}{2}$ particles, we focus on systems of up to three spins, analyzed through both analytical and numerical methods. 

Spin-$\nicefrac{1}{2}$ chains provide a natural framework for exploring fundamental aspects of quantum systems and serve as versatile models for both research and teaching. Joel, Kollmar, and Santos introduce quantum spin-$\nicefrac{1}{2}$ Heisenberg chains as prototype many-body systems, analyzing their spectrum, symmetries, and dynamics through eigenvalue and eigenstate analysis, with an emphasis on time evolution and reproducibility through openly available computer codes~\cite{Joel13}. In the context of quantum computing, Candela employs spin-$\nicefrac{1}{2}$ chains in self-contained computational projects designed for junior-level quantum mechanics students, where learners simulate quantum computers and gain hands-on experience with algorithms such as Grover’s and Shor’s~\cite{Candela15}. Beyond computation, entanglement---a central feature of quantum mechanics---is often introduced through spin systems; for instance, Schroeder demonstrates how entangled spatial wave functions can be incorporated early in quantum courses to deepen conceptual understanding, address common misconceptions, and prepare students for advanced topics such as Bell’s theorem and quantum information science~\cite{Schroeder17}. Complementing these approaches, L\'{o}pez-Incera and Dür present a classroom game that “gamifies” entanglement, engaging students as particles and scientists to investigate quantum states, measurement, decoherence, and even modern applications such as quantum cryptography~\cite{Lopez-Incera19}. Gubin and Santos study the signatures of chaotic quantum systems, with accompanying programs provided online~\cite{Gubin12}.

In this hands-on course the spins interact via the Heisenberg exchange and are subject to Zeeman coupling in external magnetic fields. We provide \texttt{python} modules\cite{code} to reproduce results and support further exploration.

This hands-on course is structured as follows: Section~\ref{sec:spins} begins with an introduction to spin states and spin operators, starting with a single spin (\ref{sec:one-spin}), progressing to two spins (\ref{sec:two-spins}), and culminating with results for three spins (\ref{sec:three-spins}). Entanglement is first introduced in Section~\ref{sec:two-spins}. In Section~\ref{sec:hamiltonian}, we present the Hamiltonian, which includes the coupling to a magnetic field (\ref{sec:hamiltonian-mag}) and spin-spin interactions (\ref{sec:hamiltonian-spin-spin}). Following this, we methodically introduce the spin dynamics (\ref{sec:spin-dynamics}) of one, two, and three spins, to analyze the transfer of entanglement (spin correlations) between spins (\ref{sec:sd3}). The Appendix points to additional states and interactions that could merit further consideration.

The course is intended for undergraduate students in their early semesters who possess a basic understanding of quantum theory---namely, familiarity with the fundamental postulates of quantum mechanics, the concepts of state vectors and operators, the Schrödinger equation, and the description of spin-$\nicefrac{1}{2}$ systems---as presented, for example, in Ref.~\onlinecite{Susskind2014}. Students are encouraged to work through the exercises presented in the following Sections. These include both analytical calculations and numerical simulations; solutions can be found in Appendix~\ref{sec:solutions}.

\section{Systems of spins}
\label{sec:spins}
In this Section, we introduce the fundamental mathematics of spin systems, beginning with a single spin and progressively extending the discussion to encompass up to three spins.

\subsection{A single spin}
\label{sec:one-spin}
If a spin or a qubit is measured, the result is either `up' or `down', in which `up' and `down' are determined by the orientation of the measurement apparatus~\cite{Susskind2014}. The Hilbert space of a single spin is thus two-dimensional. 

An often chosen orthonormal basis for this Hilbert space are the eigenstates $\ket{u}$ and $\ket{d}$ of the spin operator $\operator{\sigma}_{z}$, the latter being relevant for measurement outcomes for `spin along the $z$-direction': 
\begin{subequations}
\begin{align}
    \operator{\sigma}_{z} \ket{u} & = + \ket{u}, & \text{`up'},
    \\
    \operator{\sigma}_{z} \ket{d} & = - \ket{d}, & \text{`down'}.
\end{align}    
\end{subequations}
Here, the measurement of `up' is identified with the eigenvalue $+1$ of $\operator{\sigma}_{z}$, that of `down' by the eigenvalue $-1$. The eigenstates are orthonormal, that is
\begin{subequations}
    \begin{align}
        \braket{u | u} & = 1,
        \\
        \braket{u | d} & = 0,
        \\
        \braket{d | d} & = 1.
    \end{align}
\end{subequations}

Usually states and spin operators are \emph{represented} by vectors and hermitian matrices, respectively. For the eigenstates we write 
\begin{subequations}
\begin{align}
    \ket{u} & \longleftrightarrow \begin{pmatrix} 1 \\ 0 \end{pmatrix},
    \\
    \ket{d} & \longleftrightarrow \begin{pmatrix} 0 \\ 1 \end{pmatrix},
\end{align}
\end{subequations}
and for the spin operators one obtains with this choice the Pauli matrices
\begin{subequations}
\begin{align}
    \operator{\sigma}_{x} & \longleftrightarrow \begin{pmatrix} 0 & 1 \\ 1 & 0 \end{pmatrix},
    \\
    \operator{\sigma}_{y} & \longleftrightarrow \begin{pmatrix} 0 & -\cone \\ \cone & 0 \end{pmatrix},
    \\
    \operator{\sigma}_{z} & \longleftrightarrow \begin{pmatrix} 1 &0 \\ 0 & -1 \end{pmatrix}.
\end{align}    
\end{subequations}
The representations are derived, for example, in Ref.~\onlinecite{Susskind2014}.  For completeness, we introduce the identity operator $\operator{\sigma}_{1}$,
\begin{subequations}
    \begin{align}
        \operator{\sigma}_{1} \ket{u} & = \ket{u},
        \\
        \operator{\sigma}_{1} \ket{d} & = \ket{d},
    \end{align}
\end{subequations}
which is represented by the $2 \times 2$ identity matrix,
\begin{align}
    \operator{\sigma}_{1} & \longleftrightarrow \begin{pmatrix} 1 & 0 \\ 0 & 1 \end{pmatrix}.
\end{align}

The spin operators obey
\begin{subequations}
    \begin{align}
        \operator{\sigma}_{\mu}^{2} & = \operator{\sigma}_{1}, & \mu = x, y, z, \label{eq:SEVsquared}
        \\
        [\operator{\sigma}_{\mu}, \operator{\sigma}_{\nu}] & = 2\, \cone\,   \operator{\sigma}_{\tau}, &  \mu, \nu, \tau = x, y, z\ \text{cyclic},
    \end{align}
    \label{eq:commutator}
\end{subequations}
with the commutator $[\operator{a}, \operator{b}] \equiv \operator{a} \operator{b} - \operator{b} \operator{a}$.

\begin{exercise}[Properties of spin operators]
    Prove that the matrix representations of the spin operators fulfill Equation~\eqref{eq:commutator}.
\end{exercise}

Any quantum state $\ket{\psi}$ of a single spin can be represented in the chosen basis,
\begin{subequations}
\begin{align}
   \ket{\psi} & = \alpha_{u} \ket{u} + \alpha_{d} \ket{d} \longleftrightarrow \begin{pmatrix}
       \alpha_{u} \\ \alpha_{d} 
   \end{pmatrix},
   \\
   \bra{\psi} & = \alpha_{u}^{\star} \bra{u} + \alpha_{d}^{\star} \bra{d} \longleftrightarrow \begin{pmatrix}
       \alpha_{u}^{\star} & \alpha_{d}^{\star} 
   \end{pmatrix}.
\end{align}    
\end{subequations}
$\alpha_{u}$ and $\alpha_{d}$ are probability amplitudes (complex numbers), which obey the normalization condition
\begin{align}
 \braket{\psi | \psi} & \longleftrightarrow  \begin{pmatrix}
       \alpha_{u}^{\star} & \alpha_{d}^{\star} 
   \end{pmatrix}
   \begin{pmatrix}
       \alpha_{u} \\ \alpha_{d} 
   \end{pmatrix}
    = |\alpha_{u}|^{2} + |\alpha_{d}|^{2} = 1.
\end{align}
A single spin state is uniquely specified by two parameters: although the real and imaginary components of the two complex probability amplitudes $\alpha_{u}$ and $\alpha_{d}$ initially yield four real parameters, the total number is reduced by one due to normalization and by another due to the phase arbitrariness, leaving two independent real parameters.

Given a state $\ket{\psi}$, the probability $P_{u}$ for measuring `up' and respectively $P_{d}$ for measuring `down' are given by Born's rule, 
\begin{subequations}
\begin{align}
  P_{u} & = |\alpha_{u}|^{2},
  \\
  P_{d} & = |\alpha_{d}|^{2}. 
\end{align}    
\end{subequations}
Since each measurement of $\sigma_{z}$ yields either the eigenvalue $+1$ for $\ket{u}$ or $-1$ for $\ket{d}$, the respective expectation value reads 
\begin{align}
    \braket{\sigma_{z}}  & = (+1) P_{u} + (-1) P_{d} = |\alpha_{u}|^{2} - |\alpha_{d}|^{2}
    \label{eq:SEV1}
\end{align}
(sum over `measured value times its probability'). Equivalently, a spin expectation value $\braket{\sigma_{\mu}}$ can be calculated by the expression
\begin{align}
    \braket{\sigma_{\mu}} & = \braket{\psi | \operator{\sigma}_{\mu} | \psi}, \quad \mu = x, y, z.
\end{align}

\subsection{Two spins}
\label{sec:two-spins}
To understand the physics of a two-spin system, we begin by building on what we know from the single-spin case (Section~\ref{sec:one-spin}). The challenge is to extend our framework without losing key quantum features such as superposition, entanglement, and the non-commutative nature of quantum operators [see Equation~\eqref{eq:commutator}].

This extension is made possible through the Kronecker product (also known as the tensor product). In quantum mechanics, the state of a composite system---like two spins---is not simply a pair of single-spin states. Instead, it is described by the Kronecker product of the individual states. This construction enables the representation of entangled states, which cannot be written as a product of individual spin states. By using the Kronecker product, we ensure that the mathematical and physical structure of quantum mechanics remains intact when scaling from a single spin to a multi-spin system.

The Hilbert space of a two-spin system is four-dimensional, and using the basis introduced for a single spin, its basis states are $\ket{uu}$, $\ket{ud}$, $\ket{du}$, and $\ket{dd}$, where the first entry refers to spin~$1$ and the second entry to spin~$2$.

Each basis state is given by a Kronecker product,
\begin{align}
    \ket{ab} & = \ket{a} \otimes \ket{b}, \quad a, b = u, d.
\end{align}
For example, $\ket{ud}$ is thus represented as
\begin{align}
    \ket{ud} & \longleftrightarrow 
    \begin{pmatrix}
        1 \\ 0 
    \end{pmatrix}
    \otimes
    \begin{pmatrix}
        0 \\ 1 
    \end{pmatrix}
    = 
    \begin{pmatrix}
        1  \cdot   \begin{pmatrix}
        0 \\ 1 
    \end{pmatrix}
 \\ 0  \cdot   \begin{pmatrix}
        0 \\ 1 
    \end{pmatrix}
    \end{pmatrix}
    = 
    \begin{pmatrix}
        0 \\ 1 \\ 0 \\ 0 
    \end{pmatrix}.
\end{align}

\begin{exercise}[Representations of two-spin basis states]
    Calculate using the Kronecker product the vector representations of the other basis states $\ket{uu}$,  $\ket{du}$, and $\ket{dd}$. 
\end{exercise}

Any two-spin state can be expressed in that basis,
\begin{align}
    \ket{\psi} & = \alpha_{uu} \ket{uu} + \alpha_{ud} \ket{ud} + \alpha_{du} \ket{du} + \alpha_{dd} \ket{dd}.
    \label{eq:2spin-state}
\end{align}
Such a state is uniquely specified by six real parameters.

\begin{exercise}[Number of real parameters]
The quantum state in Eq.~\eqref{eq:2spin-state} is written in terms of four complex probability amplitudes (e.g., $\alpha_{uu}$, etc.), each of which consists of a real and an imaginary part. This would suggest that the state is described by a total of \emph{eight real parameters}.
\begin{enumerate}
    \item Which fundamental properties of quantum states lead to a reduction in the number of independent parameters?
    
    \item Explain why \emph{six real parameters} are sufficient to fully characterize the state given in Eq.~\eqref{eq:2spin-state}.
\end{enumerate}
\end{exercise}

A product state is the Kronecker product of two single-spin states,
\begin{align}
    \ket{P} & = \ket{\psi^{(1)}} \otimes \ket{\psi^{(2)}}.
\end{align}
With the two single-spin states
\begin{subequations}
\begin{align}
    \ket{\psi^{1}} & = \alpha_{u}^{(1)} \ket{u} +  \alpha_{d}^{(1)} \ket{u},
    \\
    \ket{\psi^{2}} & = \alpha_{u}^{(2)} \ket{u} +  \alpha_{d}^{(2)} \ket{u},
\end{align}    
\end{subequations}
it follows that
\begin{align}
    \ket{P} & = \alpha_{u}^{(1)}  \alpha_{u}^{(2)} \ket{uu} +  \alpha_{u}^{(1)}  \alpha_{d}^{(2)} \ket{ud} 
     +  \alpha_{d}^{(1)}  \alpha_{u}^{(2)} \ket{du} +  \alpha_{d}^{(1)}  \alpha_{d}^{(2)} \ket{dd}. 
     \label{eq:productstate}
\end{align}
Such a state is specified by four real parameters, two for each single-spin state. Hence, product states form a subset of all states in the Hilbert space.

Kronecker products allow to calculate two-spin operators,
\begin{align}
    \operator{\sigma}_{\mu \nu} & =  \operator{\sigma}_{\mu} \otimes  \operator{\sigma}_{\nu}, \quad \mu, \nu = 1, x, y, z,
\end{align}
in which $\operator{\sigma}_{\mu}$ refers to the spin~$1$ and $\operator{\sigma}_{\nu}$ to spin~$2$. An example is 
\begin{align}
    \operator{\sigma}_{xz} & \longleftrightarrow
    \begin{pmatrix}
        0 & 1 \\
        1 & 0
    \end{pmatrix}
    \otimes
    \begin{pmatrix}
        1 & 0 \\
        0 & -1
    \end{pmatrix}
    =
    \begin{pmatrix}
        0   \cdot  \begin{pmatrix}
        1 & 0 \\
        0 & -1
    \end{pmatrix}
 & 1 \cdot    \begin{pmatrix}
        1 & 0 \\
        0 & -1
    \end{pmatrix}
\\
        1 \cdot    \begin{pmatrix}
        1 & 0 \\
        0 & -1
    \end{pmatrix}
 & 0  \cdot   \begin{pmatrix}
        1 & 0 \\
        0 & -1
    \end{pmatrix}
    \end{pmatrix}
    =
    \begin{pmatrix}
        0 & 0 & 1 & 0 \\
        0 & 0 & 0 & -1 \\
        1 & 0 & 0 & 0 \\
        0 & -1 & 0 & 0
    \end{pmatrix}.
\end{align}

\begin{exercise}[Representations of two-spin operator]
    Calculate using the Kronecker product the matrix representation of a two-spin operator, say $\operator{\sigma}_{xy}$.
\end{exercise}

Expectation values for measuring the $\mu$-th spin component of \emph{one} spin are calculated by setting the identity operator as spin operator for the other spin. Explicitly, we have
\begin{align}
    \braket{\sigma_{\mu 1}} &  = \braket{\psi| \operator{\sigma}_{\mu 1} | \psi}, \quad \mu = x, y ,z,
\end{align}
for measuring spin~$1$ and
\begin{align}
    \braket{\sigma_{1 \mu}} & = \braket{\psi| \operator{\sigma}_{1 \mu} | \psi}, \quad \mu = x, y ,z,
\end{align}
for measuring spin~$2$.

Spin-spin correlation describes how the spin orientations of two particles are statistically related in a quantum system. This relationship is quantified by the spin-spin correlation function, typically defined as
\begin{align}
    \operatorname{Cor}(\mu, \nu) & = \braket{\sigma_{\mu \nu}} - \braket{\sigma_{\mu 1}} \braket{\sigma_{1 \nu}}, \quad \mu, \nu = x, y, z.
\end{align}
This expression represents a connected correlation function, which captures the part of the joint behavior of two observables that is not explained by their individual expectation values. If $\sigma_{\mu 1}$ and $\sigma_{1 \nu}$ are statistically independent, then $\braket{\sigma_{\mu \nu}} = \braket{\sigma_{\mu 1}} \braket{\sigma_{1 \nu}}$, and the connected correlation vanishes: $\operatorname{Cor}(\mu, \nu) = 0$. Thus, a nonzero connected correlation function indicates a genuine statistical (or quantum) correlation between the observables. In entangled quantum systems, such correlations often arise even when the individual expectation values vanish.

Connected spin-spin correlations are crucial for revealing the presence of quantum entanglement or interaction-driven behavior. For example, in the singlet state 
\begin{align}
    \ket{S} & = \frac{1}{\sqrt{2}} \left( \ket{ud} - \ket{du} \right),
    \label{eq:singlet-state}
\end{align}
which is not a product state, we have
\begin{subequations}
\begin{align}
    \braket{\sigma_{\mu 1}} & = 0,
    \\
    \braket{\sigma_{1 \mu}} & = 0
\end{align}
\label{eq:singlet-expect}
\end{subequations}
for all $\mu = x, y ,z$, so the connected correlation function simplifies to
\begin{align}
    \operatorname{Cor}(\mu, \nu) & = \braket{\sigma_{\mu \nu}}, \quad \mu, \nu = x, y, z.
\end{align}
These correlations are nonzero and isotropic, reflecting the perfect anti-correlation of the spins in all directions:
\begin{align}
    \operatorname{Cor}(\mu, \nu) & = 
    \begin{cases}
        -1 & \mu = \nu \\
        0  & \mu \not= \nu
    \end{cases}
    \quad \mu, \nu = x, y, z. 
    \label{eq:singlet-corr}
\end{align}
Such behavior is a hallmark of entanglement: the measurement outcome of one spin determines the outcome of the other, even though neither spin has a definite value on its own.

Besides the correlation function there are other measures for the degree of entanglement, for example the entanglement entropy. These measures  require the calculation of the density matrix, which is beyond the scope of the present hands-on course (for properties and usage of selected quantum-state measures see for example Ref.~\onlinecite{Ziolkowski23}).

\begin{exercise}[Properties of the singlet state]
In this exercise, we ask you to calculate properties of the singlet state  $\ket{S}$ defined in Equation~\eqref{eq:singlet-state}.
\begin{enumerate}
    \item Show that the singlet state $\ket{S}$ is \emph{not} a product state; that is, it cannot be written in the form given by Equation~\eqref{eq:productstate}.

    \item Prove that Equation~\eqref{eq:singlet-expect} holds for the case $\mu = z$.

    \item For the singlet state $\ket{S}$, compute the expectation values $\braket{\sigma_{zz}}$ and $\braket{\sigma_{xy}}$.

    \item Show that for $\ket{S}$, the correlation functions satisfy $\operatorname{Cor}(z, z) = -1$ and $\operatorname{Cor}(x, y) = 0$.
\end{enumerate}
\end{exercise}

\subsection{Three spins}
\label{sec:three-spins}
Extending the Hilbert space for two spins to three spins using the Kronecker product is straightforward. A basis for the eight-dimensional three-spin Hilbert space is thus   
\begin{align}
    \ket{a b c} & = \ket{a} \otimes \ket{b} \otimes \ket{c}
\end{align}
for each combination of $a$, $b$, and $c$, with $a = u, d$ for spin~1, $b = u, d$ for spin~2, and $c = u, d$ for spin~3. In vector representation this yields for example 
\begin{align}
    \ket{udd} & \longleftrightarrow 
    \begin{pmatrix}
        0 \\ 0 \\ 0 \\ 1 \\ 0 \\ 0 \\ 0 \\ 0
    \end{pmatrix}.
\end{align}
Any state is then expressed in this basis,
\begin{align}
    \ket{\psi} & =  \sum_{a = u, d} \sum_{b = u, d} \sum_{c = u, d} \alpha_{abc} \ket{abc}.
    \label{eq:3spin-state}
\end{align}
The same procedure applies to the three-spin operators
\begin{align}
    \operator{\sigma}_{\mu \nu \tau} & =  \operator{\sigma}_{\mu} \otimes  \operator{\sigma}_{\nu} \otimes  \operator{\sigma}_{\tau}, \quad \mu, \nu, \tau = 1, x, y, z.
\end{align}
The Kronecker product allows to generate systems of more spins, but for the hands-on course of this paper three spins suffice.

\section{Hamilton operator}
\label{sec:hamiltonian}
The spin states are coupled to the environment by a uniform magnetic field (with Hamiltonian $\operator{H}_{\mathrm{Z}}$); moreover, they interact pairwise with each other (with Hamiltonian $\operator{H}_{\mathrm{H}}$). The complete Hamilton operator
\begin{align}
    \operator{H} & = \operator{H}_{\mathrm{Z}} + \operator{H}_{\mathrm{H}}
    \label{eq:Hamiltonian}
\end{align}
is the sum of these two terms.

\subsection{Coupling to a magnetic field}
\label{sec:hamiltonian-mag}
We consider a uniform magnetic induction $\vec{B} = (B_{x}, B_{y}, B_{z})$ which is coupled via a Zeeman term to the individual spins, 
\begin{align}
    \operator{H}_{\mathrm{Z}} & = \frac{g \, \mu_{\mathrm{B}}}{\hbar} \vec{B} \cdot \operator{\vec{S}}
\end{align}
($g \approx 2$ Land\'{e} g-factor for electrons, $\mu_{\mathrm{B}}$ Bohr magneton). In what follows we combine magnetic field and the prefactors so that $\vec{B}$ is given in units of energy. With this simplification, the Zeeman Hamiltonian reads 
\begin{align}
    \operator{H}_{\mathrm{Z}} & \propto 
    \sum_{\mu = x, y, z} 
    \begin{cases}
    B_{\mu} \operator{\sigma}_{\mu} & \text{single spin} 
    \\
    B_{\mu} \left( \operator{\sigma}_{\mu 1} 
    + \operator{\sigma}_{1 \mu } 
    \right) &  \text{two spins}
    \\
    B_{\mu} \left( \operator{\sigma}_{\mu 1 1} 
    + \operator{\sigma}_{1 \mu 1} 
    + \operator{\sigma}_{1 1 \mu} 
    \right) &  \text{three spins}
    \end{cases}.
    \label{eq:Hm}
\end{align}
For three spins, the spin operators account for the first spin ($\operator{\sigma}_{\mu 1 1}$), the second spin ($\operator{\sigma}_{1\mu 1}$), and the third spin ($ \operator{\sigma}_{1 1 \mu}$), respectively. 

\subsection{Spin-spin interaction}
\label{sec:hamiltonian-spin-spin}
The Hamiltonian for an interaction of two spins, as has been introduced by Heisenberg \cite{Heisenberg1928,Auerbach1994,Schwabl2004}, reads
\begin{align}
    \operator{H}_{\mathrm{H}} & = -J  \sum_{\mu = x, y , z}  \operator{\sigma}_{\mu 1} \,\operator{\sigma}_{1 \mu},
\end{align}
with the interaction strength $J$ (Ref.~\onlinecite{e-e-interaction}) given in units of energy. Note that the spin operators, respectively the associated matrices, commute with each other, which implies that the $\mu$-th component of the individual spins can be measured with certainty, according to Heisenberg's uncertainty principle~\cite{Heisenberg1927}. We recall that a singlet state $\ket{S}$ is an eigenstate of the Heisenberg Hamiltonian $\operator{H}_{\mathrm{H}}$ for the respective spin pair.

Gathering all pairwise interactions for three spins yields
\begin{align}
    \operator{H}_{\mathrm{H}} & = -J  \sum_{\mu = x, y , z} 
    \left(
    \operator{\sigma}_{\mu 1 1} \, \operator{\sigma}_{1 \mu 1}
    + 
    \operator{\sigma}_{1 \mu 1} \, \operator{\sigma}_{1 1 \mu }
    +
    \operator{\sigma}_{1 1 \mu} \, \operator{\sigma}_{\mu 1 1}
    \right).
    \label{eq:HH}
\end{align}
The operator products in the parentheses account for spin pairs $1 \leftrightarrow 2$,  $2 \leftrightarrow 3$, and  $3 \leftrightarrow 1$ (from left to right). In matrix representation these become matrix multiplications.

\section{Spin dynamics}
\label{sec:spin-dynamics}
The dynamics of the spins is determined by the time-dependent Schrödinger equation 
\begin{align}
    \cone \hbar \frac{\partial}{\partial t} \ket{\psi} & = \operator{H} \ket{\psi}.
    \label{eq:TSE}
\end{align}
This initial value problem is solved numerically in the \code{python} code (Section~\ref{sec:python-code}). In order to evolve the dynamics, we proceed as follows, here exemplified for three spins:
\begin{enumerate}
    \item Prepare a three-spin initial state $\ket{\psi}$, equation~\eqref{eq:3spin-state}. \label{item:prepare}
    \item Setup the Hamiltonian $\operator{H}$, equation~\eqref{eq:Hamiltonian}.
    \item Solve numerically the time-dependent Schrödinger equation~\eqref{eq:TSE} in a given interval $t \in [0, t_{\mathrm{max}}]$, starting with the initial state prepared in step~\ref{item:prepare}. For each time $t$, spin expectation values and correlations are computed and written to disk. \label{item:simulate}
    \item Analyze data computed in step~\ref{item:simulate} (post-processing).
\end{enumerate}

\begin{exercise}[Spin in a Magnetic Field]
Consider the dynamics of a single spin described by the state (in vector representation)
\begin{align}
    \begin{pmatrix}
        \alpha_{u}(t) \\ \alpha_{d}(t)
    \end{pmatrix},
\end{align}
evolving under the influence of a magnetic field directed along the $z$-axis (confer Equation~\eqref{eq:Hm}). The time-dependent Schrödinger equation is then given by
\begin{align}
    \cone \hbar \frac{\partial}{\partial t} 
    \begin{pmatrix}
        \alpha_{u}(t) \\ \alpha_{d}(t)
    \end{pmatrix}
    = B_{0} 
    \begin{pmatrix}
        -\alpha_{d}(t) \\ \alpha_{u}(t)
    \end{pmatrix}.
\end{align}

\begin{enumerate}
    \item Solve this equation of motion for the spinor components $\alpha_u(t)$ and $\alpha_d(t)$, assuming initial values $\alpha_{u}(0)$ and $\alpha_{d}(0)$. \label{item1}
    
    \item Compute the spin polarization vector $\vec{S}(t)$, using the results from item~\ref{item1}

    \item Show that if $\vec{S}(0)$ is not aligned along the $z$-axis, then $\vec{S}(t)$ precesses around the $z$-axis.
\end{enumerate}
\end{exercise}

\section{\code{python} code}
\label{sec:python-code}
While many calculations in the preceding Sections can be done by hand, we believe that a computer program is an invaluable tool for studying the dynamics of spins and their entanglement. To this end, we developed a set of \code{python} programs \cite{code}, the structure of which is illustrated in Appendix~\ref{sec:code-structure}. We chose \code{python} as the programming language because it is widely used, well-documented, and actively maintained. NB: A different approach is to use specific computational platforms, such a quantum computers, for investigations of quantum systems~\cite{Brody21,Lancaster25}.

Concerning an integrated development environment (IDE) our choice fell on \code{Thonny} \cite{thonny-homepage,Annamaa2015a,Annamaa2015b}, which is easily installed and runs on all major operating systems. Moreover, it includes all relevant \code{python} packages, such as \code{numpy} and \code{matplotlib}, and is in our opinion well suited for beginners. We have tested the \code{python} code on other IDEs as well.

Results are visualized with the standard \code{python} package \code{matplotlib}, which also encourages students to experiment by modifying the codes themselves. An alternative approach has been proposed by Ahmed et al.~\cite{Ahmed21}, who developed \textit{Quantum Composer}, a node-based interactive tool that allows users to construct and explore quantum mechanical simulations without programming, thereby enhancing accessibility, conceptual understanding, and rapid feedback in both educational and research settings.

\section{Results}
\label{sec:results}
In all simulations the magnetic induction has a strength of $B = \unit[0.1]{meV}$ and is aligned with the $y$-axis. The Heisenberg interaction in systems with two or three spins is isotropic \cite{isotropic} with strength $J = \unit[0.2]{meV}$. All simulations cover an interval of $\unit[50]{ps}$.

\subsection{Results for a single spin}
\begin{figure}
    \centering
    \includegraphics[width = \textwidth]{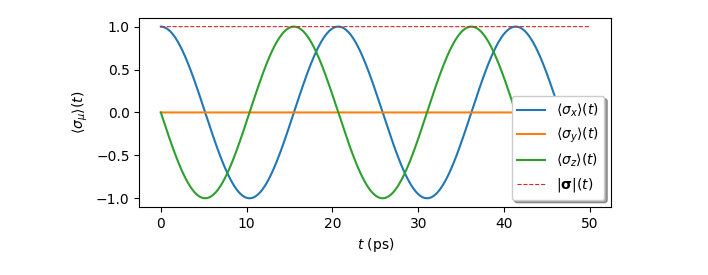}
    \caption{Precession of the spin expectation vector of a single spin in a uniform magnetic field. The three Cartesian components $\braket{\sigma_{\mu}}$, $\mu = x. y, z$, of the spin expectation vector $\vec{\sigma}$ are distinguished by line style and color, as indicated. The spin is initially in state~$\ket{r}$ (Equation~\ref{eq:right}), and the magnetic field $\vec{B}$ is along the $y$-axis. Figure generated with \code{plot-spinpol1.py}.}
    \label{fig:spin1-precession}
\end{figure}

Using \code{spindynamics1.py}, we compute the dynamics of a single spin 
in the initial state
\begin{align}
    \ket{r} & = \frac{1}{\sqrt{2}} (\ket{u} + \ket{d})
    \label{eq:right}
\end{align}
in a uniform magnetic field along the $y$-axis. Initially, at $t = 0$, the spin expectation vector 
\begin{align}
    \vec{\sigma}(t) & \equiv \begin{pmatrix}
        \braket{\sigma_{x}}(t) \\
        \braket{\sigma_{y}}(t) \\
        \braket{\sigma_{z}}(t)
    \end{pmatrix}
\end{align}
[see Equation~\eqref{eq:SEV1}] points in the $x$-direction and subsequently precesses about the $y$-axis (Fig.~\ref{fig:spin1-precession}; $\braket{\sigma_{y}} = 0$ at all times~$t$). 

The Larmor frequency, which defines the rate of precession, is directly proportional to the magnitude of the magnetic field $\vec{B}$, with $f \propto |\vec{B}|$. The precession period typically falls within the picosecond range, a common timescale for local magnetic moments in ferromagnetic materials.

In the simulations for two and three spins discussed in the next Sections, the coupling to the magnetic field is always switched on, so that the precession of the spin expectation vectors is ubiquitous. It may appear modified due to the Heisenberg interaction.

\begin{exercise}[Precession of a single spin in a magnetic field]
\begin{enumerate}
    \item Setup a simulation for a single spin in a magnetic field. Choose as initial state 
\begin{align}
    \ket{l} & = \frac{1}{\sqrt{2}} (\ket{u} - \ket{d})
    \label{eq:state-l}
\end{align}
and a magnetic field along the $z$ direction, e.g. $\vec{B} = (0, 0, 0.1)~\unit{meV}$. 

\item Show that $\ket{l}$ is an eigenstate of $\operator{\sigma}_{x}$.

\item Plot the time dependence of the spin polarization vector using \code{plot-spinpol1.py} and \code{animate-spinpol1.py}. Does the period of the precession increase with the strength of the magnetic field?
\end{enumerate}
\end{exercise}

\subsection{Results for a pair of spins}
\begin{figure}
    \centering
    \includegraphics[width = 1.0 \textwidth]{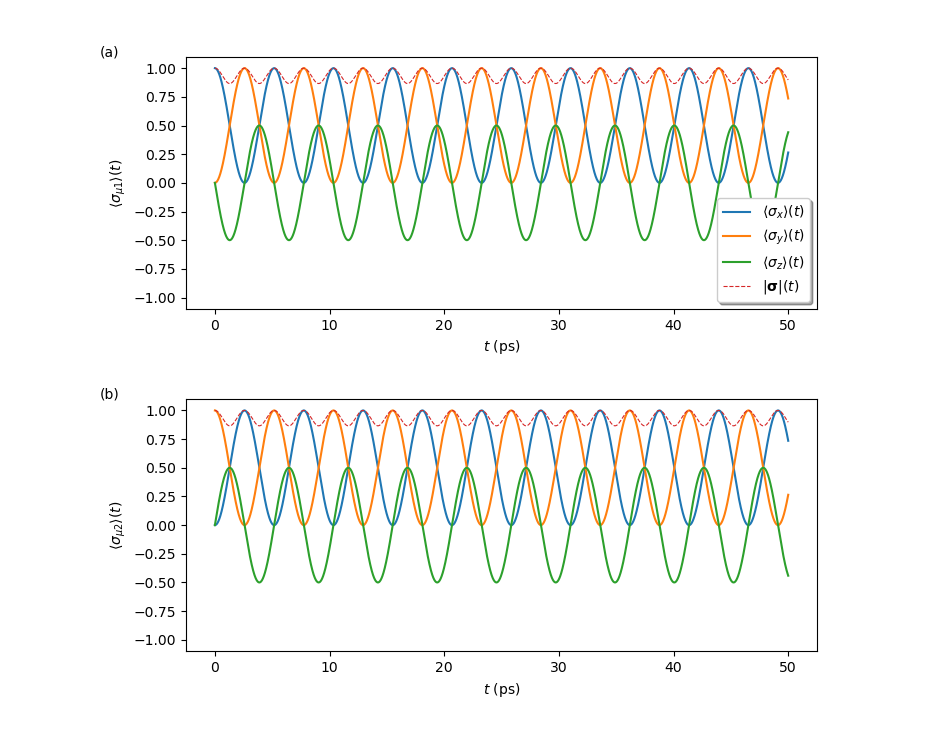}
    \caption{Dynamics of a pair of spins in the product state $\ket{l} \otimes \ket{i}$. The system evolves under the Heisenberg Hamiltonian (no magnetic field). (a) Components of the spin polarization vector of spin~$1$. (b) As (a), but for spin~$2$. The three Cartesian components $\braket{\sigma_{\mu}}$,  $\mu = x, y, z$, of the respective spin expectation vector are distinguished by line style and color, as indicated. Figure generated with \code{plot-spinpol2.py}.}
    \label{fig:spin2-PEV}
\end{figure}

\begin{figure}
    \centering
    \includegraphics[width = 1.0 \textwidth]{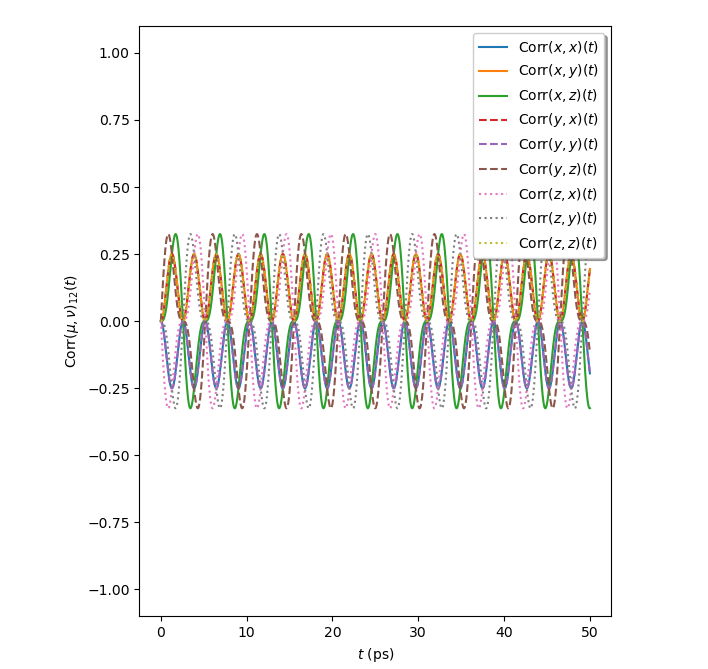}
    \caption{Correlation of a pair of spins in the product state $\ket{l} \otimes \ket{i}$, using the same parameters as for Figure~\ref{fig:spin2-PEV}; see legend.  Figure generated with \code{plot-corr.py}.}
    \label{fig:corr2-PEV}
\end{figure}

We begin with the product state $\ket{P} = \ket{l} \otimes \ket{i}$, where $\ket{l}$ is defined in Eq.~\eqref{eq:state-l} and  
\begin{align}
    \ket{i} = \frac{1}{\sqrt{2}} (\ket{u} + \cone \ket{d}).
    \label{eq:in}
\end{align}
The system evolves under a Heisenberg Hamiltonian with no external magnetic field. Notably, we observe precession of the spin polarization vectors, as illustrated in Figure~\ref{fig:spin2-PEV}. Additionally, the degree of spin polarization remains below unity, indicating a loss of information. This reduction suggests the emergence of entanglement, as the initially uncorrelated spins become correlated through the Heisenberg interaction. Indeed, the spin-spin correlation is nonzero, as shown in Figure~\ref{fig:corr2-PEV}. Moreover, we find that when the degree of spin polarization reaches a minimum, the spin-spin correlation attains its maximum (in absolute value), highlighting the interplay between local spin polarization and quantum correlations. In summary, these results demonstrate that the Heisenberg interaction can dynamically generate correlations and entanglement in a system of initially uncorrelated spins.

We now turn to the dynamics of an initially entangled state, that is the singlet state $\ket{S}$ defined in Equation~\eqref{eq:singlet-state}. However, the singlet state is an eigenstate of both the Zeeman Hamiltonian $\operator{H}_{\mathrm{Z}}$ and the Heisenberg Hamiltonian $\operator{H}_{\mathrm{H}}$; the spin expectation values and correlations hence remain constant over time, as described by Eq.~\eqref{eq:TSE}. To introduce dynamics into the system, we must therefore consider a superposition of the singlet state $\ket{S}$ and a product state $\ket{P}$. Moreover, the Hamiltonian now includes not only the coupling of individual spins to the magnetic field $\vec{B}$ but also the Heisenberg interaction between the two spins. 

\begin{figure}
    \centering
    \includegraphics[width = 1.0 \textwidth]{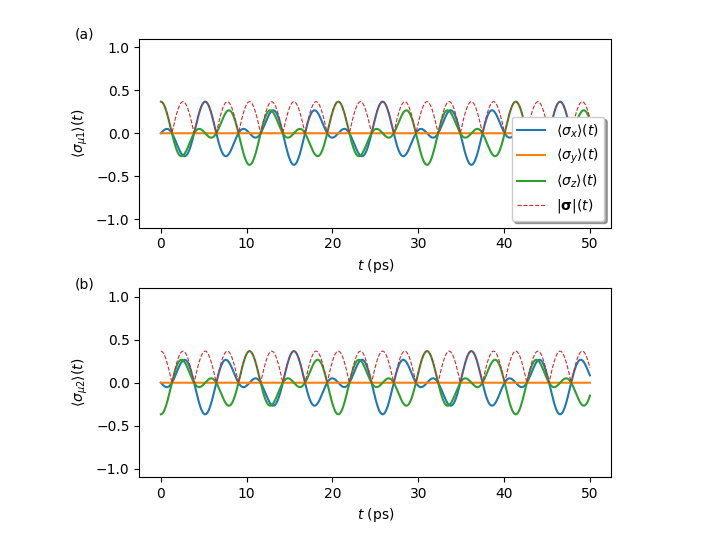}
    \caption{Dynamics of the pair of spins initially in the state givn by Equation~\eqref{eq:superposition-state}. (a) Components of the spin polarization vector of spin~$1$. (b) As (a), but for spin~$2$. The three Cartesian components $\braket{\sigma_{\mu}}$,  $\mu = x, y, z$, of the respective spin expectation vector are distinguished by line style and color, as indicated. Figure generated with \code{plot-spinpol2.py}.}
    \label{fig:spin2-SEV}
\end{figure}

\begin{figure}
    \centering
    \includegraphics[width = \textwidth]{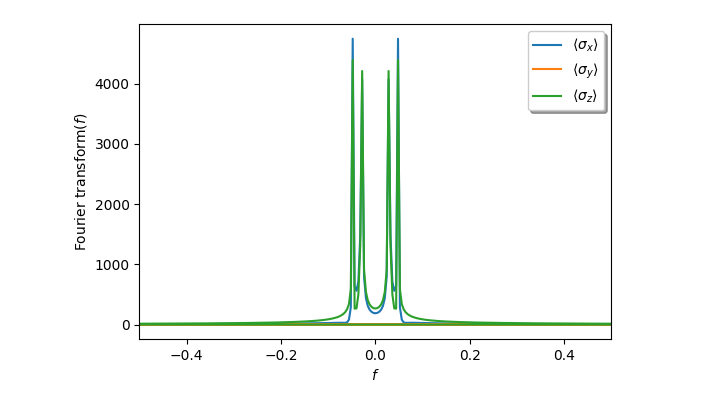}
    \caption{FFT spectrum of the spin expectation values $\braket{\sigma_{\mu 1}}$, $\mu = x, y, z$, of spin~1 (distinguished by color, as indicated. The spectrum of spin~2 is identical to that of spin~1. The initial state is defined in Equation~\ref{eq:superposition-state}. Figure generated with \code{plot-spin-fft.py}.}
    \label{fig:spin2-fft}
\end{figure}

Here we present results for the initial state
\begin{align}
    \ket{\Psi} \propto \frac{3}{4} \ket{S} + \frac{1}{4} \ket{P},
    \label{eq:superposition-state}
\end{align}
in which the product state $\ket{P}$ is specified by $\alpha_{u}^{(1)} = \alpha_{d}^{(2)} = 1$ and $\alpha_{d}^{(1)} = \alpha_{u}^{(2)} = 0$ [confer Eq.~\eqref{eq:productstate}]. As expected, this superposition results in time-dependent spin expectation values and new correlations 
(Fig.~\ref{fig:spin2-SEV}). Instead of a `pure sinusoidal' behavior of the spin expectation values for a single spin (Fig.~\ref{fig:spin1-precession}), the Heisenberg interaction results in a mixture of two superimposed oscillations with well-defined frequencies, as is evident from the FFT spectra depicted in Figure~\ref{fig:spin2-fft}. As for a single spin the expectation values, $\braket{\sigma_{y1}}$ and $\braket{\sigma_{1y}}$ vanish.

\begin{figure}
    \centering
    \includegraphics[width = \textwidth]{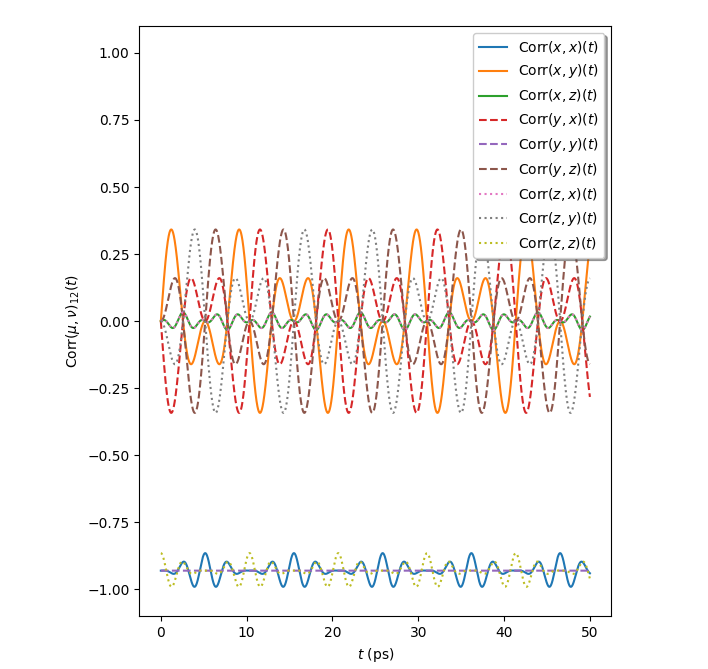}
    \caption{Correlation between the Cartesian components of the two spins (distinguished by color, as indicated). Figure generated with \code{plot-corr.py}.}
    \label{fig:spin2-corr}
\end{figure}

The correlations for identical Cartesian components are now oscillating with a mean value close to, but different from $-1$ (Fig.~\ref{fig:spin2-corr}). Recall that for the singlet state $\operatorname{Cor}(\mu, \mu) = -1$, $\mu = x, y, z$. Moreover, the mixing  of the product state with the singlet state results in nonzero correlations for different Cartesian components ($\mu \not = \nu$), which oscillate in a sinusoidal manner about $0$.

As the correlations evolve over time, the system undergoes entanglement dynamics driven by the Heisenberg interaction. This initial observation suggests that the interaction can transfer entanglement to subsystems that were initially not entangled with the rest of the system.

In summary, rather than exhibiting a pure sinusoidal behavior as observed with a single spin, the Heisenberg interaction introduces a combination of two oscillations with distinct frequencies. Unlike the single spin scenario where $\braket{\sigma_{y1}}$ and $\braket{\sigma_{1y}}$ are zero, the Heisenberg interaction modifies correlations between identical Cartesian components. Additionally, the mixture of the product state with the singlet state results in nonzero correlations for different Cartesian components.

\begin{exercise}[Dynamics of two spins] \label{exercise:two-spin}
Simulate the dynamics of two spins, one spin initially in state $\ket{l}$, defined in Eq.~\eqref{eq:state-l}, the other in state $\ket{i}$, defined in Eq.~\eqref{eq:in}.
\begin{enumerate}
    \item In what direction does the spin polarization vector of state $\ket{i}$ point?

    \item Calculate the probability amplitudes of the initial two-spin state $\ket{P} = \ket{l} \otimes \ket{i}$. See Eq.~\eqref{eq:productstate}.
    
    \item Perform a simulation for $\vec{B} = (0, 0, 0.1)~\unit{meV}$ and $J = 0$. 

    \item Perform a simulation for $\vec{B} = 0$ and $J = \unit[0.2]{meV}$.

    \item Perform a simulation for $\vec{B} = (0, 0, 0.1)~\unit{meV}$ and $J = \unit[0.2]{meV}$.
\end{enumerate}
Analyze the spin polarizations, in particular the periods. In what case(s) does a beating pattern show up? Argue, why a beating pattern could show up. You might want to use FFT to extract the frequencies (\code{plot-spin-fft.py}).
\end{exercise}

\begin{exercise}[Dynamics of the singlet state.] \label{exercise:singlet}
Repeat exercise~\ref{exercise:two-spin} but for the singlet state $\ket{S}$ defined in Eq.~\eqref{eq:singlet-state}. 

\begin{enumerate}
    \item There should be no dynamics in the system; specifically, the spin polarization of each individual spin remains zero at all times.
    Why does this phenomenon occur?

    Hint: Consider the time-dependent Schrödinger equation~\eqref{eq:TSE} for a quantum state that is an eigenstate of the Hamiltonian operator~$\operator{H}$. In this case, the equation can be solved analytically. Does the time evolution introduce only a global phase factor in the solution? 

    \item Do the spin-spin correlations depend on time? Use \code{plot-corr.py} for plotting these data.
\end{enumerate}
\end{exercise}

\begin{exercise}[Dynamics of a superposition state]
    Now perform a simulation of the superposition of a product state and the singlet state. Compare the dynamics with that of a product state, as in the exercise~\ref{exercise:two-spin}, and the singlet state, as in the exercise~\ref{exercise:singlet}.
\end{exercise}

\subsection{Results for three spins}
\label{sec:sd3}
In what follows we consider the three-spin state 
\begin{align}
    \ket{\psi} & = \frac{1}{\sqrt{2}} \left( \ket{udu} - \ket{duu} \right).
    \label{eq:3spinstate}
\end{align}
built of a singlet state $\ket{S}$ for spins~1 and~2 as well as spin~3 in the $\ket{u}$ state. This combination of an entangled spin pair $\ket{S}$ with a third spin $\ket{u}$ that is not entangled with the singlet pair lends itself for studying the transfer of entanglement between spins. Although the spatial layout of the spins plays no role in our analysis, it is convenient to visualize them as equally spaced along a linear chain. Then the Heisenberg interaction couples the central spin~2 with the `edge' spins 1 and 3:
\begin{align}
    \operator{H}_{\mathrm{H}} & = -J  \sum_{\mu = x, y , z} 
    \left(
    \operator{\sigma}_{\mu 1 1} \, \operator{\sigma}_{1 \mu 1}
    + 
    \operator{\sigma}_{1 \mu 1} \, \operator{\sigma}_{1 1 \mu }
    \right).
    \label{eq:HHchain}
\end{align}

\begin{figure}
    \centering
    \includegraphics[width = 0.75 \textwidth]{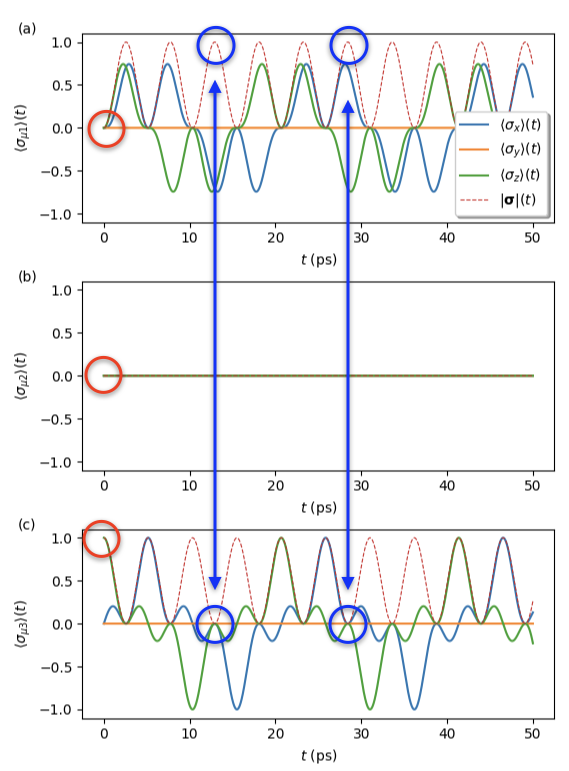}
    \caption{Dynamics of three spins, where initially (at $t = \unit[0]{ps}$)  spin $1$ (a) and spin $2$ (b) form a singlet state and spin~$3$ (c) is in an up state; see Eq.~\eqref{eq:3spinstate}. Cartesian components $\braket{\sigma_{\mu i}}(t)$ and modulus $| \vec{\sigma}|(t)$ of the spin expectation values are distinguished by color and line style (as indicated). Red circles mark the initial spin polarizations: all three components of spins 1 and and 2 are zero, thus indicating the entangled singlet state, whereas spin 3 is in the $\ket{u}$ state ($\braket{\sigma_{z}} = 1$). Blue circles and arrows mark representative times at which spins 2 and 3 are entangled (zero spin polarization), but spin 1 is neither entangled with spin 2 nor with spin 3. Figure generated with \code{plot-spinpol3.py}, but annotated subsequently to highlight important features.}
    \label{fig:spin3-precession}
\end{figure}

This initial setup produces the spin expectation values $\braket{\sigma_{\mu}^{1}} = \braket{\sigma_{\mu}^{2}} = 0$ at $t = \unit[0]{ps}$, as expected for an entangled state. For spin $3$ we have $\braket{\sigma_{x}^{3}} = \braket{\sigma_{y}^{3}} = 0$ and $\braket{\sigma_{z}^{3}} = 1$, which indicates zero entanglement with the singlet pair  (Fig.~\ref{fig:spin3-precession}). At first glance surprising, the spin expectation values of spin~$2$ remains zero at all times~$t$, whereas those of the other two spins oscillate at $t > \unit[0]{ps}$. If there were no Heisenberg coupling of  spin~$2$ with spin~$3$, spin~$3$ would show the usual precession about $\vec{B}$ (cf.\ Fig.~\ref{fig:spin1-precession}); this precession is hence modified by the Heisenberg interaction with spin~$2$.

All spin expectation values $\braket{\sigma_{\mu 3}}$ for spin~$3$ vanish, for instance, at $t \approx \unit[13]{ps}$ (Fig.~\ref{fig:spin3-precession}c). This can only occur if spin~$3$ becomes entangled with the other two spins. At this moment, $\braket{\sigma_{x 1}}$ and $\braket{\sigma_{z 1}}$ are nonzero, each with a value of $-\sqrt{2}/2$, summing to a maximum spin polarization of $|\vec{\sigma}| = 1$ (with $\braket{\sigma_{y 1}} = 0$; Fig.~\ref{fig:spin3-precession}b). Additionally, the spin polarization of spin~2 remains zero throughout (Fig.~\ref{fig:spin3-precession}b). This indicates entanglement between spins~2 and 3, demonstrating that the initial entanglement between spins~1 and 2 has been transferred to spins~2 and 3. This transfer of entanglement occurs periodically.

The observation that all spin expectation values for spin~$2$ remain zero at all times can be confirmed through analytical calculations. This result is an artifact of the specific choices of $J$ and $B$ (here: $\unit[0.2]{meV}$ and $\unit[0.1]{meV}$, respectively): other parameter combinations produce oscillating spin expectation values for spin~$2$. Additionally, it implies that correlations can be transferred between spins coupled via the Heisenberg interaction. Further simulations (not presented here) reinforce this idea, indicating that even in the absence of a magnetic field, the spin expectation values of spin~$3$ would still oscillate. We encourage the reader to explore this phenomenon further.

\begin{figure}
    \centering
    \includegraphics[width = 0.75 \textwidth]{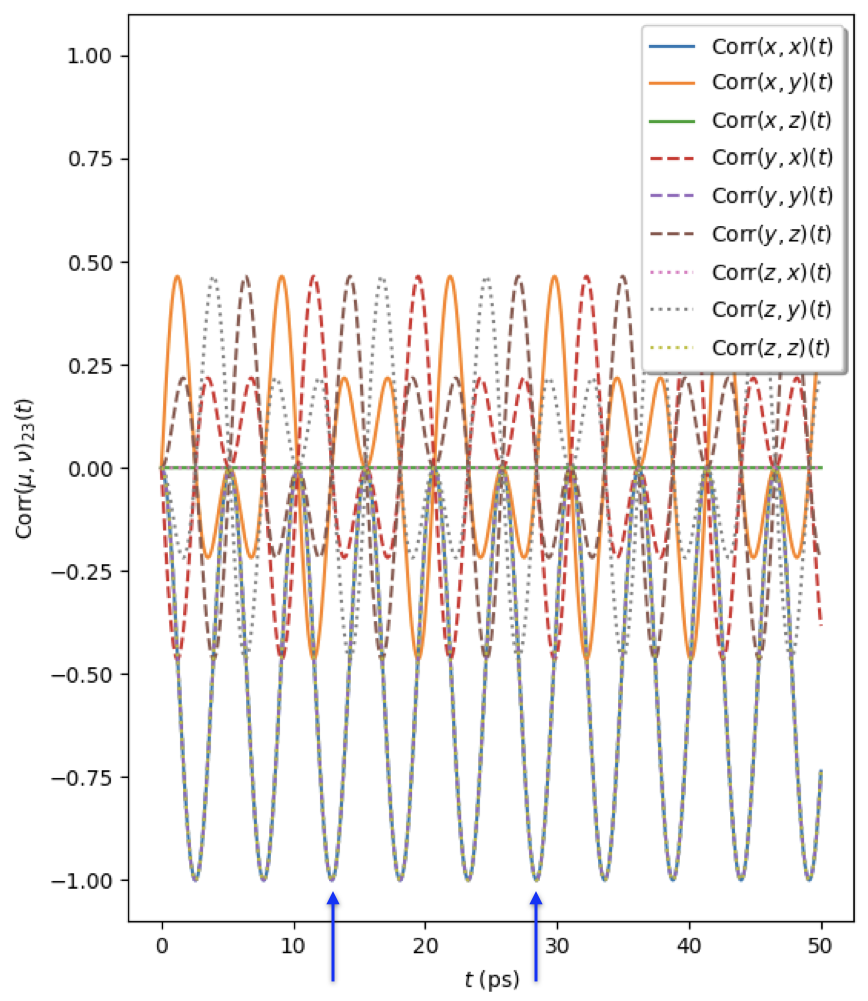}
    \caption{Dynamics of three spins, as in Fig.~\ref{fig:spin3-precession}. The spectra represent correlations $\operatorname{Corr}(\mu, \nu)(t)$ ($\mu, \nu = x, y, z)$ between spin~$2$ and spin~$3$. Blue arrows mark representative times at which $\operatorname{Corr}(\mu, \mu) = -1$ ($\mu = x, y, z)$, indicating complete anticorrelation. Figure generated with \code{plot-corr.py}, but annotated subsequently to highlight important features.}
    \label{fig:spin3-corr}
\end{figure}

The transfer of entanglement is further evidenced by inspecting correlations. As an example, we show the correlations between spins~$2$ and $3$ (Fig.~\ref{fig:spin3-corr}), which are not entangled at $t = 0$. As discussed before, these spins become entangled at $t \approx \unit[13]{ps}$. This finding is nicely confirmed by correlations $\operatorname{Corr}(\mu, \mu) = -1$ ($\mu = x, y, z$) and $\operatorname{Corr}(\mu, \nu) = 0$ (for $\mu \not= \nu$), as for a singlet state. Hence we conclude that now spins~$2$ and $3$ form a singlet state, while spin~$1$ is not entangled with this spin pair. The latter finding is corroborated by inspecting the correlations of spin~$1$ with spin~$2$ and spin~$3$, respectively (not shown here). The oscillatory nature of these correlations demonstrates that entanglement can be transferred back and forth along a spin chain through the Heisenberg interaction.

\begin{exercise}[Speed of entanglement transfer]
    In Figure~\ref{fig:spin3-corr} we have identified times at which $\operatorname{Corr}(\mu, \mu) = -1$ ($\mu = x, y, z)$, indicating complete anticorrelation.

    How do these characteristic times depend on the coupling strength $J$ in the Heisenberg Hamiltonian? These times define a speed of entanglement transfer. Perform respective simulations and analyze how this speed varies with $J$.
\end{exercise}

In summary, we have confirmed that entanglement can be transferred between spins through our simulations. The oscillatory behavior observed in the correlations between different spin pairs, such as spins~$1$ and $3$, demonstrates that entanglement is indeed transferable along a spin chain via the Heisenberg interaction.

\section{Conclusions and outlook}
\label{sec:conclusions}
Our simulations underscore the dynamic nature of quantum correlations in spin systems and highlight the ability of the Heisenberg interaction to facilitate entanglement transfer between spins.

Looking ahead, the \code{python} code modules developed in this study provide a foundation for exploring systems with more than three spins. The flexible nature of the code allows for various boundary conditions, such as implementing a ring of spins. While our current study focuses on systems where the spatial arrangement of spins is irrelevant, future work could extend this to two- or even three-dimensional configurations, offering a richer understanding of spatial effects on spin dynamics. Additionally, exact diagonalization of the spin Hamiltonian will be crucial for analyzing more complex systems and accurately characterizing their properties. An exciting avenue for future research includes investigating the transfer of entanglement in large samples, possibly illuminating decoherence~\cite{Schlosshauer2007} and the quantum-mechanical measurement process.

\appendix
\section{Triplet states}
\label{sec:triplet}
For the sake of completeness, we recall the three triplet states
\begin{subequations}
    \begin{align}
        \ket{T_{1}} & = \frac{1}{\sqrt{2}} \left( \ket{ud} + \ket{du} \right),
        \\
        \ket{T_{2}} & = \frac{1}{\sqrt{2}} \left( \ket{uu} + \ket{dd} \right),
        \\
        \ket{T_{3}} & = \frac{1}{\sqrt{2}} \left( \ket{uu} - \ket{dd} \right),        
    \end{align}
    \label{eq:state-triplets}
\end{subequations}
which are also entangled and, together with the singlet state $\ket{S}$, form another basis for the two-spin Hilbert space.

\section{Dzyaloshinskii-Moriya interaction}
As a side note, spin-orbit coupling of electrons results in an additional spin-spin interaction, that is the Dzyaloshinskii-Moriya interaction~\cite{Dzyaloshinskii1958,Morya1960} for two spins,
\begin{align}
    \operator{H}_{\mathrm{D}} & \propto \vec{D} \cdot \left( \vec{\operator{\sigma}}^{(1)} \times \vec{\operator{\sigma}}^{(2)} \right),
    \label{eq:HD}
\end{align}
with
\begin{align}
    \vec{\operator{\sigma}}^{(1)} & \equiv \begin{pmatrix}
        \operator{\sigma}_{x 1} \\
        \operator{\sigma}_{y 1} \\
        \operator{\sigma}_{z 1}
    \end{pmatrix}
\end{align}
and
\begin{align}
    \vec{\operator{\sigma}}^{(2)} & \equiv \begin{pmatrix}
        \operator{\sigma}_{1 x} \\
        \operator{\sigma}_{1 y} \\
        \operator{\sigma}_{ 1z}
    \end{pmatrix}.
\end{align}
This interaction is not considered in this hands-on course but may be taken into account in the \code{python} code (Section~\ref{sec:python-code}).

\section{Structure of the \code{python} code}
\label{sec:code-structure}
\begin{figure}
    \centering
    \includegraphics[width=0.95\textwidth]{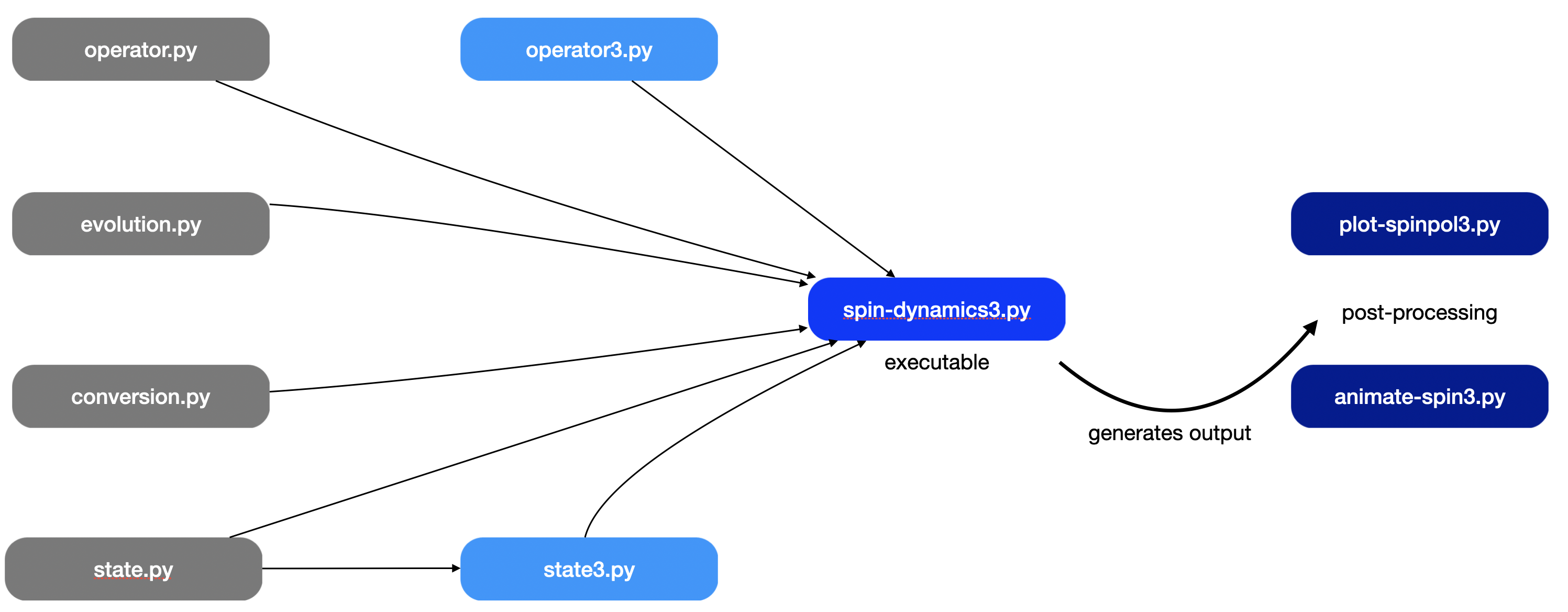}
    \caption{Dependencies of the \code{python} modules for simulations of three spins. For details, see text.}
    \label{fig:code-structure}
\end{figure}

The \code{python} program \code{spin-dynamics3.py} performs the time evolution briefly described in Section~\ref{sec:spin-dynamics}; its dependencies on other \code{python} modules is sketched in Fig.~\ref{fig:code-structure}. In its first lines, the user has to specify the probability amplitudes $\alpha_{abc}$ of the initial state in Equation~\eqref{eq:3spin-state}, the vector of the magnetic induction $\vec{B}$ in Equation~\eqref{eq:Hm}, and the Heisenberg parameters coupling strength $J$ in Equation~\eqref{eq:HH}. Optionally the Dzyloshinski-Moriya parameters in Equation~\eqref{eq:HD} may be specified. Moreover, the time interval and the integrator for the Schrödinger equation~\eqref{eq:TSE} have to be specified. We provide with \code{spin-dynamics1.py} and \code{spin-dynamics2.py} similar programs for one and two spins.

In the module \code{evolution.py}, we implemented several propagators for solving the time-dependent Schrödinger equation~\eqref{eq:TSE}, including the Euler method, the Heun method, and two midpoint methods. This variety allows for a comparison of the numerical accuracy of these methods as an additional feature.

The \code{state3.py} and \code{operator3.py} modules provide functions for generating and manipulating spin states and operators, respectively. \code{conversion.py} delivers functions for unit conversions, e.g., for time and energy.

\code{spin-dynamics3.py} as well as the other spin-dynamics modules generate output data files, which are subsequently used for analysis (Fig.~\ref{fig:code-structure}). The evolution of spin expectation values is visualized by \code{plot-spinpol3.py} and animated using \code{animate-spin3.py}, while \code{plot-corr.py} is used to plot spin correlations. Additionally, the fast Fourier transform (FFT) of spin and correlation data can be analyzed with \code{plot-spin-fft.py} and \code{plot-corr-fft.py}, respectively.

The code is free software: it may be copied, distributed and/or modified under the terms of the GNU General Public License as published by the Free Software Foundation, either version 3 of the License or any later version \cite{GPLv3}.

\section{Solutions to the exercises}
\label{sec:solutions}

\subsection{Solution to exercise 1}
\label{sec:solution1}

We are to prove two properties of the Pauli spin operators:
\begin{enumerate}
    \item each Pauli matrix squares to the identity,
    \begin{align*}
    \sigma_\mu^2 = \sigma_{1} \quad \text{for } \mu \in \{x, y, z\}
    \end{align*}
    and 
    \item the commutation relation
    \begin{align*}
    [\sigma_\mu, \sigma_\nu] = 2 \cone \sigma_\tau, \quad \text{for } \mu, \nu, \tau \in  \{x, y, z\} \ \text{cyclic}.
    \end{align*}
\end{enumerate}

The Pauli matrices are
\begin{align*}
\sigma_x = 
\begin{pmatrix}
0 & 1 \\
1 & 0
\end{pmatrix}, \quad
\sigma_y = 
\begin{pmatrix}
0 & -\cone \\
i & 0
\end{pmatrix}, \quad
\sigma_z = 
\begin{pmatrix}
1 & 0 \\
0 & -1
\end{pmatrix}.
\end{align*}
Furthermore, 
\begin{align*}
\sigma_1 = 
\begin{pmatrix}
1 & 0 \\
0 & 1
\end{pmatrix}.
\end{align*}

\subsubsection{Squared Pauli matrices}
We now show that each Pauli matrix squared gives the identity matrix.

\paragraph*{$\sigma_x^2$:}
\begin{align*}
\sigma_x^2 =
\begin{pmatrix}
0 & 1 \\
1 & 0
\end{pmatrix}
\begin{pmatrix}
0 & 1 \\
1 & 0
\end{pmatrix}
=
\begin{pmatrix}
1 & 0 \\
0 & 1
\end{pmatrix}
= \sigma_{1}.
\end{align*}

\paragraph*{$\sigma_y^2$:}
\begin{align*}
\sigma_y^2 =
\begin{pmatrix}
0 & -\cone \\
\cone & 0
\end{pmatrix}
\begin{pmatrix}
0 & -\cone \\
\cone & 0
\end{pmatrix}
=
\begin{pmatrix}
1 & 0 \\
0 & 1
\end{pmatrix}
= \sigma_{1}.
\end{align*}

\paragraph*{$\sigma_z^2$:}
\begin{align*}
\sigma_z^2 =
\begin{pmatrix}
1 & 0 \\
0 & -1
\end{pmatrix}
\begin{pmatrix}
1 & 0 \\
0 & -1
\end{pmatrix}
=
\begin{pmatrix}
1 & 0 \\
0 & 1
\end{pmatrix}
= \sigma_{1}.
\end{align*}

Hence, for all $\mu \in \{x, y, z\}$
\begin{align*}
\sigma_\mu^2 & = \sigma_{1}
\end{align*}
holds.

\subsubsection{Commutation relations}
Let us verify $[ \sigma_x, \sigma_y ] = 2\cone \sigma_z$:
\begin{align*}
\sigma_x \sigma_y &=
\begin{pmatrix}
0 & 1 \\
1 & 0
\end{pmatrix}
\begin{pmatrix}
0 & -\cone \\
i & 0
\end{pmatrix}
=
\begin{pmatrix}
\cone & 0 \\
0 & -\cone
\end{pmatrix}
= \cone \sigma_z,
\\
\sigma_y \sigma_x &=
\begin{pmatrix}
0 & -\cone \\
\cone & 0
\end{pmatrix}
\begin{pmatrix}
0 & 1 \\
1 & 0
\end{pmatrix}
=
\begin{pmatrix}
-\cone & 0 \\
0 & \cone
\end{pmatrix}
= -\cone \sigma_z,
\end{align*}
hence
\begin{align*}
[\sigma_x, \sigma_y] &= \sigma_x \sigma_y - \sigma_y \sigma_x = 2 \cone \sigma_z.
\end{align*}

Similarly, one proves
\begin{align*}
[\sigma_y, \sigma_z] = 2\cone \sigma_x,
\\
[\sigma_z, \sigma_x] = 2 \cone \sigma_y.
\end{align*}
Thus
\begin{align*}
[\sigma_\mu, \sigma_\nu] = 2\cone \sigma_\tau, \quad \mu, \nu, \tau = x, y ,z\ \text{cyclic}.
\end{align*}

\subsection{Solution to exercise 2}
\label{sec:solution2}
We use the Kronecker (tensor) product to compute the vector representations of the two-spin basis states.

The single-spin basis vectors are represented by
\begin{align*}
\ket{u} & = 
\begin{pmatrix}
1 \\
0
\end{pmatrix},
\\
\ket{d} & = 
\begin{pmatrix}
0 \\
1
\end{pmatrix}.
\end{align*}

Let us compute the Kronecker products for the two-spin states.

\paragraph*{$\ket{uu}$:}
\begin{align*}
\ket{uu} = \ket{u} \otimes \ket{u} =
\begin{pmatrix}
1 \\
0
\end{pmatrix}
\otimes
\begin{pmatrix}
1 \\
0
\end{pmatrix}
=
\begin{pmatrix}
1 \cdot \begin{pmatrix}
1 \\
0
\end{pmatrix}
 \\
0 \cdot \begin{pmatrix}
1 \\
0
\end{pmatrix}
\end{pmatrix}
=
\begin{pmatrix}
1 \\
0 \\
0 \\
0
\end{pmatrix}.
\end{align*}

\paragraph*{$\ket{du}$:}
\begin{align*}
\ket{du} = \ket{d} \otimes \ket{u} =
\begin{pmatrix}
0 \\
1
\end{pmatrix}
\otimes
\begin{pmatrix}
1 \\
0
\end{pmatrix}
=
\begin{pmatrix}
0 \cdot \begin{pmatrix}
1 \\
0
\end{pmatrix}
 \\
1 \cdot \begin{pmatrix}
1 \\
0
\end{pmatrix}
\end{pmatrix}
=
\begin{pmatrix}
0 \\
0 \\
1 \\
0
\end{pmatrix}.
\end{align*}

\paragraph*{$\ket{dd}$:}
\begin{align*}
\ket{dd} = \ket{d} \otimes \ket{d} =
\begin{pmatrix}
0 \\
1
\end{pmatrix}
\otimes
\begin{pmatrix}
0 \\
1
\end{pmatrix}
=
\begin{pmatrix}
0 \cdot \begin{pmatrix}
0 \\
1
\end{pmatrix}
 \\
1 \cdot \begin{pmatrix}
0 \\
1
\end{pmatrix}
\end{pmatrix}
=
\begin{pmatrix}
0 \\
0 \\
0 \\
1
\end{pmatrix}.
\end{align*}

So the full two-spin computational basis in vector form is
\begin{align*}
\ket{uu} = 
\begin{pmatrix}
1 \\ 0 \\ 0 \\ 0
\end{pmatrix}, \quad
\ket{ud} = 
\begin{pmatrix}
0 \\ 1 \\ 0 \\ 0
\end{pmatrix}, \quad
\ket{du} = 
\begin{pmatrix}
0 \\ 0 \\ 1 \\ 0
\end{pmatrix}, \quad
\ket{dd} = 
\begin{pmatrix}
0 \\ 0 \\ 0 \\ 1
\end{pmatrix}.
\end{align*}

\subsection{Solution to exercise 3}
\label{sec:solution3}
We are given that a general two-spin quantum state is written as
\begin{align*}
\ket{\psi} = \alpha_{uu} \ket{uu} + \alpha_{ud} \ket{ud} + \alpha_{du} \ket{du} + \alpha_{dd} \ket{dd},
\end{align*}
where each coefficient $\alpha_{ij}$ is a complex number.

\paragraph*{What reduces the number of independent parameters?}
Initially, there are four complex coefficients, each with a real and imaginary part:
\begin{align*}
4 \text{ complex numbers} \Rightarrow 8 \text{ real parameters}.
\end{align*}
However, quantum states are not uniquely defined by their components alone; they must satisfy the following constraints:
\begin{itemize}
    \item Normalization: The state vector must have unit norm:
    \begin{align*}
    \braket{\psi | \psi} = |\alpha_{uu}|^2 + |\alpha_{ud}|^2 + |\alpha_{du}|^2 + |\alpha_{dd}|^2 = 1.
    \end{align*}
    This condition removes \emph{one real parameter}.
    
    \item Global phase invariance: A global phase factor $e^{\cone\phi}$ multiplying $\ket{\psi}$ has no physical consequence:
    \begin{align*}
    \ket{\psi} & \sim e^{\cone\phi} \ket{\psi}.
    \end{align*}
    This means that both states have identical expectation values for all observables, rendering them indistinguishable. Therefore, an overall complex phase does not affect measurement outcomes and is not physically observable. This removes \emph{one additional real parameter}.
\end{itemize}

\paragraph*{Why six real parameters are sufficient?}
Starting from eight real parameters:
\begin{align*}
8 \text{ (from 4 complex numbers)} & - 1 \text{ (normalization)} - 1 \text{ (global phase)} \\
& = 6 \text{ real parameters}.
\end{align*}
Hence, only six real parameters are needed to fully specify the physical (measurable) content of a general two-spin state.

\subsection{Solution to exercise 4}
\label{sec:solution4}
We are asked to compute the matrix representation of the two-spin operator $\operator{\sigma}_{xy}$, which is defined as
\begin{align*}
\operator{\sigma}_{xy} = \operator{\sigma}_x \otimes \operator{\sigma}_y.
\end{align*}

We begin with the matrix representations of the Pauli matrices,
\begin{align*}
\sigma_x & = 
\begin{pmatrix}
0 & 1 \\
1 & 0
\end{pmatrix},
\\
\sigma_y & = 
\begin{pmatrix}
0 & -\cone \\
\cone & 0
\end{pmatrix}.
\end{align*}
The Kronecker product $\sigma_x \otimes \sigma_y$ is computed as follows:
\begin{align*}
\sigma_x \otimes \sigma_y & =
\begin{pmatrix}
0 & 1 \\
1 & 0
\end{pmatrix}
\otimes
\begin{pmatrix}
0 & -\cone\\
\cone & 0
\end{pmatrix}
=
\begin{pmatrix}
0 \cdot \begin{pmatrix}
0 & -\cone\\
\cone & 0
\end{pmatrix}
 & 1 \cdot \begin{pmatrix}
0 & -\cone\\
\cone & 0
\end{pmatrix}
 \\
1 \cdot \begin{pmatrix}
0 & -\cone\\
\cone & 0
\end{pmatrix}
 & 0 \cdot \begin{pmatrix}
0 & -\cone\\
\cone & 0
\end{pmatrix}
\end{pmatrix}
\\
& =
\begin{pmatrix}
0 & 0 & 0 & -\cone \\
0 & 0 & \cone & 0 \\
0 & -\cone & 0 & 0 \\
\cone & 0 & 0 & 0
\end{pmatrix}.
\end{align*}
Therefore, the matrix representation of the operator $\operator{\sigma}_{xy}$ is
\begin{align*}
\sigma_x \otimes \sigma_y =
\begin{pmatrix}
0 & 0 & 0 & -\cone \\
0 & 0 & \cone & 0 \\
0 & -\cone & 0 & 0 \\
\cone & 0 & 0 & 0
\end{pmatrix}.
\end{align*}

\subsection{Solution to exercise 5}
\label{sec:solution5}
The singlet state is defined as
\begin{align*}
\ket{S} = \frac{1}{\sqrt{2}} (\ket{ud} - \ket{du}).
\end{align*}

\paragraph*{Singlet state is not a product state:}
Suppose the singlet state were a product state:
\begin{align*}
\ket{S} = \ket{\psi^{(1)}} \otimes \ket{\psi^{(2)}} = ( \alpha_{u}^{(1)} \ket{u} + \alpha_{d}^{(1)} \ket{d}) \otimes (\alpha_{u}^{(2)} \ket{u} +  \alpha_{d}^{(2)}\ket{d}).
\end{align*}
Expanding the product yields
\begin{align*}
\alpha_{u}^{(1)} \alpha_{u}^{(2)} \ket{uu} + \alpha_{u}^{(1)} \alpha_{d}^{(2)} \ket{ud} + \alpha_{d}^{(1)}\alpha_{u}^{(2)} \ket{du} + \alpha_{d}^{(1)}\alpha_{d}^{(2)} \ket{dd}.
\end{align*}
This product state must match the singlet state:
\begin{align*}
\ket{S} = \frac{1}{\sqrt{2}} (\ket{ud} - \ket{du}).
\end{align*}

Matching terms gives:
\begin{itemize}
    \item The coefficients of $\ket{uu}$ and $\ket{dd}$ must be zero: $\alpha_{u}^{(1)} \alpha_{u}^{(2)} = 0$ and  $\alpha_{d}^{(1)} \alpha_{d}^{(2)} = 0$.

    \item The coefficient of $\ket{ud}$ must be $\frac{1}{\sqrt{2}}$: $\alpha_{u}^{(1)} \alpha_{d}^{(2)} = \frac{1}{\sqrt{2}}$.

    \item The coefficient of $\ket{du}$ must be $-\frac{1}{\sqrt{2}}$: $\alpha_{d}^{(1)} \alpha_{u}^{(2)} = -\frac{1}{\sqrt{2}}$.
\end{itemize}
From $\alpha_{u}^{(1)} \alpha_{u}^{(2)} = 0$ follows that either $\alpha_{u}^{(1)} = 0$ or $\alpha_{u}^{(2)} = 0$. And from $\alpha_{d}^{(1)} \alpha_{d}^{(2)}  = 0$ follows that either $\alpha_{d}^{(1)} = 0$ or $\alpha_{d}^{(2)} = 0$.

If $\alpha_{u}^{(1)} = 0$, then $\alpha_{u}^{(1)} \alpha_{d}^{(2)} = 0$, contradicting $\alpha_{u}^{(1)} \alpha_{d}^{(2)} = \frac{1}{\sqrt{2}}$.  And 
if $\alpha_{u}^{(2)} = 0$, then $\alpha_{d}^{(2)}\alpha_{u}^{(1)}  = 0$, contradicting $\alpha_{d}^{(2)}\alpha_{u}^{(1)}  = -\frac{1}{\sqrt{2}}$.

Similar contradictions arise for all combinations.

Therefore, no choice of $\alpha_{u}^{(1)}$, $\alpha_{d}^{(1)}$, $\alpha_{u}^{(2)}$, and $\alpha_{d}^{(2)}$ can reproduce the singlet state as a product state.  Hence, we conclude that the singlet state is not a product state.

\paragraph*{Show that Eq.~\eqref{eq:singlet-expect} holds for $\mu = z$:}
Equation~\eqref{eq:singlet-state} states
\begin{align*}
    \braket{\sigma_{\mu 1}} & = 0,
    \\
    \braket{\sigma_{1 \mu}} & = 0
\end{align*}
for all $\mu, \nu = x, y ,z$. We check for $\mu = z$ and use
\begin{align*}
\operator{\sigma}_z \ket{u} = \ket{u}, \quad \operator{\sigma}_z \ket{d} = -\ket{d}.
\end{align*}
Then
\begin{align*}
\operator{\sigma}_{z1} \ket{ud} & = \operator{\sigma}_z \ket{u} \otimes \operator{\sigma}_1 \ket{d} = (+1)\ket{u} \otimes 1 \ket{d} = \ket{ud}
\end{align*}
and 
\begin{align*}
\operator{\sigma}_{z1} \ket{du} & = \operator{\sigma}_1 \ket{d} \otimes \operator{\sigma}_z \ket{u} = (-1) \ket{d} \otimes \ket{d} = - \ket{du}.
\end{align*}
With this and using the orthogonality of the basis states we have
\begin{align*}
    \braket{\sigma_{z1}} & = \braket{S | \operator{\sigma}_{z1} | S}
    \\
    & = \frac{1}{2} \left( \bra{ud} - \bra{du} \right) \operator{\sigma}_{z1} \left( \ket{ud} - \ket{du}\right) 
    \\
    & = \frac{1}{2} \left( \bra{ud} - \bra{du} \right)  \left( \ket{ud} + \ket{du}\right) 
    \\
    & = \frac{1}{2} \left( \braket{ud | ud} + \braket{ud | du}  - \braket{du | ud} - \braket{du | du} \right) 
    \\
    & = \frac{1}{2} \left( 1 + 0 - 0 - 1\right) 
    \\
    & = 0.
\end{align*}

Similarly, one proves $\braket{\sigma_{1 \mu}} = 0$.

\paragraph*{Compute $\braket{\sigma_{zz}}$ and $\braket{\sigma_{xy}}$:}
In analogy to the preceding calculation, we calculate
\begin{align*}
    \braket{\sigma_{zz}} & = \braket{S | \operator{\sigma}_{zz} | S}
    \\
    & = \frac{1}{2} \left( \bra{ud} - \bra{du} \right) \operator{\sigma}_{zz} \left( \ket{ud} - \ket{du}\right) 
    \\
    & = \frac{1}{2} \left( \bra{ud} - \bra{du} \right)  \left( -\ket{ud} + \ket{du}\right) 
    \\
    & = \frac{1}{2} \left( -\braket{ud | ud} + \braket{ud | du}  + \braket{du | ud} - \braket{du | du} \right) 
    \\
    & = \frac{1}{2} \left( -1 + 0 + 0 - 1\right) 
    \\
    & = -1.
\end{align*}

Now we compute $\braket{S | \operator{\sigma}_{xy} | S}$. Using
\begin{align*}
\operator{\sigma}_x \ket{u} = \ket{d}, \quad \operator{\sigma}_x \ket{d} = \ket{u}, \quad
\operator{\sigma}_y \ket{u} = \cone \ket{d}, \quad \operator{\sigma}_y \ket{d} = -\cone \ket{u}
\end{align*}
(which follows from the vector and matrix representations), we find
\begin{align*}
\operator{\sigma}_{xy} \ket{ud} & = \operator{\sigma}_x \ket{u} \otimes \operator{\sigma}_y \ket{d} = \ket{d} \otimes (-\cone \ket{u}) = -\cone \ket{du}
\end{align*}
and 
\begin{align*}
\operator{\sigma}_{xy} \ket{du} & = \operator{\sigma}_x \ket{d} \otimes \operator{\sigma}_y \ket{u} = \ket{u} \otimes \cone \ket{d} = \cone \ket{ud}.
\end{align*}
This gives
\begin{align*}
\operator{\sigma}_{xy} \ket{S} & = \frac{1}{\sqrt{2}} (-\cone \ket{du} - \cone \ket{ud}) = \cone \frac{1}{\sqrt{2}} (\ket{ud} + \ket{du}) = -\cone \ket{T_{1}}
\end{align*}
(the triplet state $\ket{T}_{1}$ is defined in Appendix~\ref{sec:triplet}). The action of $\operator{\sigma}_{xy}$ on $\ket{S}$ produces the triplet state $\ket{T_{1}}$ (times $-\cone$); since the latter is orthogonal to $\ket{S}$, we arrive at
\begin{align*}
    \braket{\sigma_{xy}} & = \cone \braket{S | T_{1}} = 0.
\end{align*}
For completeness we prove that $\ket{S}$ and $\ket{T_{1}}$ are indeed orthogonal:
\begin{align*}
    \braket{S | T_{1}} & = \frac{1}{2} \left( \bra{ud} - \bra{du} \right) \left( \ket{ud} + \ket{du}\right)
    \\
    & = \frac{1}{2} \left( \braket{ud | ud} + \braket{ud | du} - \braket{du | ud} - \braket{du | du} \right) 
    \\
    & = \frac{1}{2} \left( 1 + 0 - 0 - 1\right) 
    \\
    & = 0.
\end{align*}

\paragraph*{Correlations:}
Using the above results and the definition of the correlation,
\begin{align*}
    \operatorname{Cor}(\mu, \nu) & = \braket{\sigma_{\mu \nu}} - \braket{\sigma_{\mu 1}} \braket{\sigma_{1 \nu}}, \quad \mu, \nu = x, y, z,
\end{align*}
we conclude
\begin{align*}
    \operatorname{Cor}(zz) & = \braket{\sigma_{zz}} - \braket{\sigma_{z1}} \braket{\sigma_{1z}} = -1 + 0 \cdot 0 = -1
\end{align*}
and
\begin{align*}
    \operatorname{Cor}(xy) & = \braket{\sigma_{xy}} - \braket{\sigma_{x1}} \braket{\sigma_{1y}} = 0 + 0 \cdot 0 = 0.
\end{align*}

\subsection{Solution to exercise 6}
\label{sec:solution6}
\begin{enumerate}
    \item The time-dependent Schrödinger equation can be written as
    \begin{align*}
    \frac{\partial}{\partial t} 
    \begin{pmatrix}
       \alpha_{u}(t) \\
       \alpha_{d}(t)
    \end{pmatrix}
    & = - \cone \frac{B_{0}}{\hbar}
    \begin{pmatrix}
        1 & 0 \\
        0 & -1
    \end{pmatrix}
    \begin{pmatrix}
       \alpha_{u}(t) \\
       \alpha_{d}(t)
    \end{pmatrix}
    \end{align*}
  There are several methods to solve this system of coupled differential equations: one can eliminate a variable by differentiating one of the equations with respect to time, apply an exponential \textit{ansatz}, or reformulate and solve the system in matrix form. In this analysis, we adopt the matrix approach.

The above equation represents a set of two coupled, homogeneous, first-order linear differential equations, which implies the existence of two independent solutions. Introducing 
    \begin{align*}
        \omega & = \frac{B_{0}}{\hbar},
    \end{align*}
    the eigenvalues and eigenvectors of the matrix
    \begin{align*}
    - \cone \omega 
    \begin{pmatrix}
        1 & 0 \\
        0 & -1
    \end{pmatrix}
    & = - \cone \omega \sigma_{z}
    \end{align*}
    are
    \begin{align*}
        - \cone \omega & : \begin{pmatrix}
            1 \\ 0
        \end{pmatrix},
        \\
        + \cone \omega & : \begin{pmatrix}
            0 \\ 1
        \end{pmatrix}.
    \end{align*}
    One solution is thus
    \begin{align*}
        \begin{pmatrix}
            1 \\ 0 
        \end{pmatrix}
        \euler^{- \cone \omega t},
    \end{align*}
    the other 
    \begin{align*}
        \begin{pmatrix}
            0 \\ 1 
        \end{pmatrix}
        \euler^{+ \cone \omega t},
    \end{align*}
    so that the general solution is given by their superposition
    \begin{align*}
     \begin{pmatrix}
       \alpha_{u}(t) \\
       \alpha_{d}(t)
    \end{pmatrix}
    & = \alpha_{u}(0) 
        \begin{pmatrix}
            1 \\ 0 
        \end{pmatrix}
        \euler^{- \cone \omega t}
    + \alpha_{d}(0)
        \begin{pmatrix}
            0 \\ 1 
        \end{pmatrix}
        \euler^{+ \cone \omega t}.
    \end{align*}
    The exponentials tell that after a period of $T = 2 \pi / \omega$ the same state is obtained, 
    \begin{align*}
        \alpha_{u}(t + T) & = \alpha_{u}(t),
        \\
        \alpha_{d}(t + T) & = \alpha_{d}(t).
    \end{align*}
    \label{item1p}
    
    \item The components of the spin polarization vector $\vec{S}(t)$ can be expressed as
    \begin{align*}
        S_{x}(t) & = 2 \, \Re\left( \alpha_{u}(t)^{\star} \, \alpha_{d}(t) \right),
        \\
        S_{y}(t) & = 2 \, \Im\left( \alpha_{u}(t)^{\star} \, \alpha_{d}(t) \right),
        \\
        S_{z}(t) & = | \alpha_{u}(t)|^{2} - | \alpha_{d}(t)|^{2}.
    \end{align*}
    From the above result we have
    \begin{align*}
        \alpha_{u}(t) & = \alpha_{u}(0) \, \euler^{- \cone \omega t},
        \\
        \alpha_{d}(t) & = \alpha_{d}(0) \, \euler^{+ \cone \omega t}.
    \end{align*}
    With this one calculates
    \begin{align*}
        S_{x}(t) & = 2 \Re \left( \alpha_{u}^{\star}(0) \, \alpha_{d}(0) \, \euler^{+ 2 \cone \omega t} \right),
        \\
        S_{y}(t) & = 2 \Im \left( \alpha_{u}^{\star}(0) \, \alpha_{d}(0) \, \euler^{+ 2 \cone \omega t} \right),
        \\
        S_{z}(t) & = |\alpha_{u}(0)|^{2} -  |\alpha_{d}(0)|^{2}.
    \end{align*}
    The exponential terms in $S_{x}(t)$ and $S_{y}(t)$ indicate that these components repeat after a period of $\pi / \omega$, which is half of the period  of the quantum state that is discussed in item~\ref{item1p}. This behavior is characteristic for spin-$1/2$ particles. Furthermore, the final  equation shows  that $S_{z}(t)$ remains constant over time. 
    
    \item The components of the spin polarization vector $\vec{S}(t)$ indicate that it precesses about the $z$-axis with a frequency of $2 \omega$. We illustrate this behavior in more detail for two specific quantum states.
    \begin{enumerate}
        \item We set the initial amplitudes to $\alpha_{u}(0) = 1$ and $\alpha_{d}(0) = 0$ which corresponds to the spin-up state $\ket{u}$. The spin polarization vector is calculated as
        \begin{align*}
            \vec{S}(t) & = \begin{pmatrix}
                0 \\ 0 \\ 1
            \end{pmatrix}.
        \end{align*}
        There is no precession, since $\vec{S}(t)$ is aligned along the magnetic field. The same arises for the state $\ket{d}$.

        \item Now we set $\alpha_{u}(0) = 1 / \sqrt{2}$ and $\alpha_{d}(0) = 1 / \sqrt{2}$. $\vec{S}(t)$ then reads
        \begin{align*}
        \vec{S}(t) & = 
        \begin{pmatrix}
        \cos(2 \omega t),
        \\
        \sin(2 \omega t),
        \\
        0            
        \end{pmatrix}.
    \end{align*}
    This vector rotates counterclockwise in the $xy$ plane with frequency $2 \omega$. It is initially oriented in $+x$ direction ($t = 0$, so the corresponding state may be denoted $\ket{r}$ (with `r' indicating `right-pointing').
    \end{enumerate}
\end{enumerate}

\subsection{Solution to exercise 7 (item 2)}
\label{sec:solution7}
We are given the state `left',
\begin{align*}
\ket{l} = \frac{1}{\sqrt{2}} (\ket{u} - \ket{d}),
\end{align*}
and asked to show that it is an eigenstate of $\operator{\sigma}_x$.

Recall that the Pauli matrix $\sigma_x$ is defined as
\begin{align*}
\sigma_x = 
\begin{pmatrix}
0 & 1 \\
1 & 0
\end{pmatrix}.
\end{align*}
Using the vector representation of the basis states
\begin{align*}
\ket{u} = 
\begin{pmatrix}
1 \\
0
\end{pmatrix}, \quad
\ket{d} = 
\begin{pmatrix}
0 \\
1
\end{pmatrix}.
\end{align*}
Then,
\begin{align*}
\ket{l} = \frac{1}{\sqrt{2}} \left( \ket{u} - \ket{d} \right) = \frac{1}{\sqrt{2}} 
\begin{pmatrix}
1 \\
-1
\end{pmatrix}.
\end{align*}
Now apply $\operator{\sigma}_x$ to $\ket{l}$; in the vector-matrix representation this gives
\begin{align*}
\begin{pmatrix}
0 & 1 \\
1 & 0
\end{pmatrix}
\cdot
\frac{1}{\sqrt{2}}
\begin{pmatrix}
1 \\
-1
\end{pmatrix}
=
\frac{1}{\sqrt{2}}
\begin{pmatrix}
-1 \\
1
\end{pmatrix}
=
- \frac{1}{\sqrt{2}}
\begin{pmatrix}
1 \\
-1
\end{pmatrix}
= -\ket{l}.
\end{align*}
Therefore, $\ket{l}$ is an eigenstate of $\operator{\sigma}_x$ with eigenvalue $-1$:
\begin{align*}
\operator{\sigma}_x \ket{l} = -\ket{l}.
\end{align*}

\subsection{Solution to exercise 8 (items 1 and 2)}
\label{sec:solution8}
We are given two initial single-spin states,
\begin{align*}
\ket{l} = \frac{1}{\sqrt{2}} (\ket{u} - \ket{d}), \quad
\ket{i} = \frac{1}{\sqrt{2}} (\ket{u} + \cone \ket{d}).
\end{align*}

\paragraph*{Direction of the spin polarization vector of $\ket{i}$:}
The spin polarization vector of a state
\begin{align*}
\ket{\psi} = \alpha_{u} \ket{u} + \alpha_{d} \ket{d}
\end{align*}
is given by the vector of expectation values
\begin{align*}
\vec{S} = \left( \braket{\sigma_x}, \braket{\sigma_y}, \braket{\sigma_z} \right),
\end{align*}
with
\begin{align*}
\braket{\sigma_x} &= \bra{\psi} \sigma_x \ket{\psi} = \alpha_{u}^* \alpha_{d} + \alpha_{d}^* \alpha_{u}, \\
\braket{\sigma_y} &= \bra{\psi} \sigma_y \ket{\psi} = -\cone (\alpha_{u}^* \alpha_{d} - \alpha_{d}^* \alpha_{u}), \\
\braket{\sigma_z} &= \bra{\psi} \sigma_z \ket{\psi} = |\alpha_{u}|^2 - |\alpha_{d}|^2.
\end{align*}

For the state $\ket{i}$ we identify
\begin{align*}
\alpha_{u} = \frac{1}{\sqrt{2}}, \quad \alpha_{d} = \frac{\cone}{\sqrt{2}}
\end{align*}
and compute the components
\begin{align*}
\braket{\sigma_x} &= \frac{1}{2} \left( 1 \cdot \cone + (-\cone) \cdot 1 \right)  = 0, \\
\braket{\sigma_y} & = - \frac{\cone}{2} \left( 1 \cdot \cone - (-\cone) \cdot 1 \right) = 1, \\
\braket{\sigma_z} & = \frac{1}{2} - \frac{1}{2} = 0.
\end{align*}
Thus, the spin polarization vector of $\ket{i}$ is
\begin{align*}
\vec{S} = (0, 1, 0).
\end{align*}
It points in the positive $y$-direction.

\paragraph*{Probability amplitudes of the two-spin product state $\ket{P} = \ket{l} \otimes \ket{i}$:}
We compute the Kronecker product
\begin{align*}
\ket{P} &= \left( \frac{1}{\sqrt{2}}(\ket{u} - \ket{d}) \right) \otimes \left( \frac{1}{\sqrt{2}}(\ket{u} + \cone \ket{d}) \right) \\
    &= \frac{1}{2} \left(
    \ket{uu} + \cone \ket{ud} - \ket{du} - \cone \ket{dd}
    \right).
\end{align*}
Therefore, the two-spin state in the standard basis $\{\ket{uu}, \ket{ud}, \ket{du}, \ket{dd}\}$ has the following probability amplitudes:
\begin{align*}
\alpha_{uu} &= \frac{1}{2}, \\
\alpha_{ud} &= \frac{\cone}{2}, \\
\alpha_{du} &= -\frac{1}{2}, \\
\alpha_{dd} &= -\frac{\cone}{2}.
\end{align*}

\subsection{Solution to exercise 9 (item 1)}
\label{sec:solution9}
\paragraph*{Coupling to a magnetic field:}
We consider the Zeeman Hamiltonian
\begin{align*}
    \operator{H}_{\mathrm{Z}}  & = -B_{z} \left( \operator{\sigma}_{z1} + \operator{\sigma}_{1z} \right)
\end{align*}
for two spins in a uniform magnetic field along the $z$-axis. The singlet state is defined as
\begin{align*}
\ket{S} = \frac{1}{\sqrt{2}}(\ket{ud} - \ket{du}).
\end{align*}

We show that $\ket{S}$ is an eigenstate of $\operator{H}_{\mathrm{Z}}$. Recall the action of $\operator{\sigma}_{z}$ on the basis states:
\begin{align*}
\operator{\sigma}_{\mathrm{z}} \ket{u} = \ket{u}, \quad \operator{\sigma}_{\mathrm{z}} \ket{d} = -\ket{d}.
\end{align*}
Now apply the Hamiltonian to the two-spin basis states in $\ket{S}$. For $\ket{ud}$ we have
\begin{align*}
\operator{\sigma}_{z1} \ket{ud} & = \ket{ud},  \\
\operator{\sigma}_{1z} \ket{ud} & = -\ket{ud}.
\end{align*}
Hence, 
\begin{align*}
    \operator{H}_{\mathrm{Z}} \ket{ud} &= -B_{z} (1 - 1) \ket{ud} = 0.
\end{align*}
And since for $\ket{du}$
\begin{align*}
\operator{\sigma}_{z1} \ket{du} & = -\ket{du}, \\
\operator{\sigma}_{1z} \ket{du} & =  \ket{du},
\end{align*}
also
\begin{align*}
    \operator{H}_{\mathrm{Z}} \ket{du} &= -B_{z} (-1 - +) \ket{ud} = 0
\end{align*}
holds. Combing these two results, we find
\begin{align*}
\operator{H}_{\mathrm{Z}} \ket{S} &= \frac{1}{\sqrt{2}} \left( \operator{H}_{\mathrm{Z}} \ket{ud} - \operator{H}_{\mathrm{Z}} \ket{du} \right) = \frac{1}{\sqrt{2}} (0 - 0) = 0.
\end{align*}
The singlet state $\ket{S}$ is an eigenstate of $\operator{H}_{\mathrm{Z}}$ with eigenvalue $0$. Therefore, the time-dependent Schrödinger equation
\begin{align*}
    \cone \hbar \frac{\partial}{\partial t} \ket{S(t)} & = \operator{H}_{\mathrm{Z}} \ket{S(t)} = 0
\end{align*}
states that $\ket{S(t)}$ is constant. This means that all expectation values of observables, such as spin polarization and correlations, are constant in time as well.

\paragraph*{Heisenberg interaction:}
We consider the spin-spin interaction (Heisenberg interaction) 
\begin{align*}
    \operator{H}_{\mathrm{H}} & = -J  \sum_{\mu = x, y , z}  \operator{\sigma}_{\mu 1} \,\operator{\sigma}_{1 \mu}
\end{align*}
and its action on the singlet state $\ket{S}$. In the following we discuss the $\mu = z$ contribution ($\operator{\sigma}_{z 1} \,\operator{\sigma}_{1 z}$); the contributions for $\mu = x$ and $\mu = y$ give the same results.

From the matrix representation for the two-spin operators it follows
\begin{align*}
  \sigma_{z 1} \cdot \sigma_{1 z} & = \sigma_{zz}
\end{align*}
(see also Exercise~4), which acts on the two-spin basis states as
\begin{align*}
\operator{\sigma}_{zz} \ket{ud} & = -\ket{ud},
\\
\operator{\sigma}_{zz} \ket{du} &= -\ket{du}.
\end{align*}
Applying this operator to $\ket{S}$ gives
\begin{align*}
\operator{\sigma}_{zz} \ket{S}
& = \operator{\sigma}_{zz} \left( \frac{1}{\sqrt{2}}(\ket{ud} - \ket{du}) \right) \\
&= \frac{1}{\sqrt{2}}(-\ket{ud} + \ket{du}) \\
&= -\frac{1}{\sqrt{2}}(\ket{ud} - \ket{du}) \\
&= -\ket{S}.
\end{align*}
We conclude that the singlet state $\ket{S}$ is an eigenstate of $\operator{\sigma}_{zz}$ with eigenvalue $-1$. Similarly one shows that $\ket{S}$ is also an eigenstate of $\operator{\sigma}_{xx}$ and $\operator{\sigma}_{yy}$ with the same eigenvalue $-1$.

The time-dependent Schrödinger equation then becomes 
\begin{align*}
    \cone \hbar \frac{\partial}{\partial t} \ket{S(t)} & = \operator{H}_{\mathrm{H}} \ket{S(t)}
    \\
    & =  -J  \sum_{\mu = x, y , z}  \operator{\sigma}_{\mu\mu} \ket{S(t)}
    \\
    & =  +J  \sum_{\mu = x, y , z}  \ket{S(t)}
    \\
    & =  3 J  \ket{S(t)},
\end{align*}
which is solved by 
\begin{align*}
    \ket{S(t)} & = e^{-3 \cone J t / \hbar} \ket{S(0)}.
\end{align*}
Hence, the time dependence is given by a global phase., and as a result all spin expectation values are constant due to the global phase invariance.

\acknowledgments
This work is funded by the Deutsche Forschungsgemeinschaft (DFG, German Research Foundation)---Project-ID 328545488---TRR~227, project~B04.

\bibliographystyle{unsrtnat}
\bibliography{references}

\begin{thebibliography}{29}
\bibitem[iso()]{isotropic}
For quantum spins, ``isotropic'' means the Heisenberg Hamiltonian is
  SU(2)-invariant (invariant under all global spin rotations), just as in the
  classical case it is SO(3)-invariant (depending on the scalar product of the
  spin vectors).
\end{thebibliography}

\end{document}